\pdfoutput=1
\documentclass[10pt,journal,compsoc]{IEEEtran}
\usepackage{multicol}
\usepackage{amssymb}
\usepackage{amsmath}
\usepackage{pifont}
\usepackage{graphicx}
\usepackage{float}
\usepackage[utf8]{inputenc}    
\usepackage{url}
\usepackage{graphicx}
\usepackage{amssymb}
\usepackage{amsmath}
\usepackage{calrsfs}
\usepackage{pifont}
\usepackage{caption}
\usepackage{subcaption}
\usepackage{float}
\usepackage[utf8]{inputenc}    
\usepackage[table, svgnames, dvipsnames]{xcolor}
\usepackage{multirow}
\usepackage{titlesec}
\usepackage{algorithm}
\usepackage[noend]{algpseudocode}
\usepackage{bbm}
\usepackage{tikz}
\usetikzlibrary{arrows}
\newcommand*{\ident}[1]{\texttt{\small #1}}

\tikzstyle{refines} = [->,->]

\newcommand{\refi}[3]{$\begin{tikzpicture}[baseline=-0.63ex]%
\node (A) {#1};%
\node (B) [right of=A, node distance=4em] {#3};%
\draw[refines] (A) -- node[midway,above=-2pt] {#2} (B);%
\end{tikzpicture}$}
\usepackage[table, svgnames, dvipsnames]{xcolor}
\usepackage{multirow}
\usepackage{longtable}
\usepackage[flushleft]{threeparttable}
\usepackage{amsmath}
\usepackage{amssymb}
\usepackage{placeins}   
\usepackage{setspace}
\usepackage{nth}
\usepackage{nomencl}
\makenomenclature

\ifCLASSOPTIONcompsoc
  \usepackage[nocompress]{cite}
\else
  \usepackage{cite}
\fi

\begin{document}

\title{Frequent Pattern Mining in Continuous-time Temporal Networks}
\author{Ali~Jazayeri
        and~Christopher C.~Yang
\IEEEcompsocitemizethanks{\IEEEcompsocthanksitem A. Jazayeri and C. C. Yang are with College of Computing and Informatics, Drexel University (\{ali.jazayeri, chris.yang\}@drexel.edu).}
\thanks{}}

\markboth{}%
{Jazayeri \& Yang: Frequent Pattern Mining in Continuous-time Temporal Networks}

\IEEEtitleabstractindextext{%
\begin{abstract}
Networks are used as highly expressive tools in different disciplines. In recent years, the analysis and mining of temporal networks have attracted substantial attention. Frequent pattern mining is considered an essential task in the network science literature. In addition to the numerous applications, the investigation of frequent pattern mining in networks directly impacts other analytical approaches, such as clustering, quasi-clique and clique mining, and link prediction. In nearly all the algorithms proposed for frequent pattern mining in temporal networks, the networks are represented as sequences of static networks. Then, the inter- or intra-network patterns are mined. This type of representation imposes a computation-expressiveness trade-off to the mining problem. In this paper, we propose a novel representation that can preserve the temporal aspects of the network losslessly. Then, we introduce the concept of constrained interval graphs (CIGs). Next, we develop a series of algorithms for mining the complete set of frequent temporal patterns in a temporal network data set. We also consider four different definitions of isomorphism to allow noise tolerance in temporal data collection. Implementing the algorithm for three real-world data sets proves the practicality of the proposed algorithm and its capability to discover unknown patterns in various settings.
\end{abstract}

}

\maketitle

 \nomenclature{$agg_w$}{Aggregation window}
\nomenclature{$\mathbb{W}$}{Time window of temporal network}
\nomenclature{$N/T$}{Temporal network}
\nomenclature{$V$}{Set of vertices}
\nomenclature{$E$}{Set of edges}
\nomenclature{$IG$}{Interval Graph}
\nomenclature{$IT$}{Interval Tree}
\nomenclature{$s$}{Subgraph}
\nomenclature{$freq(s)$}{Frequency of subgraph $s$}
\nomenclature{$\mathcal{I}$}{Time interval}
\nomenclature{$CIG$}{Constrained Interval Graph}
\nomenclature{$DAG$}{Directed Acyclic Graph}
\nomenclature{$\mathcal{L}$}{List of edges of temporal network}
\nomenclature{$\mathcal{M}$}{Map of vertices to their associated interval trees}
\nomenclature{$DS$}{Data set of temporal networks}
\nomenclature{$DS'$}{Data set of $CIG$s associated with temopral networks}
\nomenclature{$\epsilon$}{User-defined support threshold}
\nomenclature{$cl$}{Canonical labeling}

 \printnomenclature
 
\IEEEdisplaynontitleabstractindextext
\IEEEpeerreviewmaketitle




\section{Introduction}
\label{sec:introduction}
\IEEEPARstart{N}{etworks} have been extensively adopted for modeling systems where in addition to the systems' components, the inter-component interactions may provide deeper insights into the systems' behavior. Networks, with a long history of applications \cite{Bertalanffy1972, Newman2003}, are used as highly expressive tools for system modeling in different domains \cite{Luciano2011applications}.

Among different analytical and mining techniques proposed for network research, the mining of frequent network patterns has an essential place \cite{Jiang2013survey}. The underlying idea behind this problem is that the recurring patterns observed more frequently may represent essential characteristics of the system that networks represent \cite{yoshida1994graph, BorgeltBerthold2002}. However, the implementation of network mining for identifying frequent patterns is a non-trivial and computationally costly task. The main reason is the requirement to verify graph and subgraph isomorphism in different iterations of the frequent pattern mining process. 

Furthermore, in many applications, the temporality of the systems should be included in the modeling effort. It is shown that when the time scale of the changes in the system is comparable,  using dynamic and time-varying network models can inform the identification of important components more effectively \cite{Tangetal2010centralitymetrics, Kostakos2009Temporal}. One approach is to represent the temporal aspects of the system as attributes of the vertices and edges of the corresponding network.  However, this approach might obscure some of the temporal information \cite{Kovanenetal2011Temporalmotifs, Nicosia2013Temporal}. Besides, some of the well-defined metrics and concepts in static networks, such as distance, diameter, centralities, and connectivity, have been differently defined and interpreted for temporal networks. Therefore, aggregating and representing temporal aspects of the system as static networks' characteristics might adversely impact the derived insight \cite{Tangetal2010TemporalDistance, Kempeetal2002Connectivity, Pan2011temporal, Holme2012Temporal, Holme2015colloquium}. 

 \begin{figure*}[htbp]
\centering
    \includegraphics[width=0.7\linewidth]{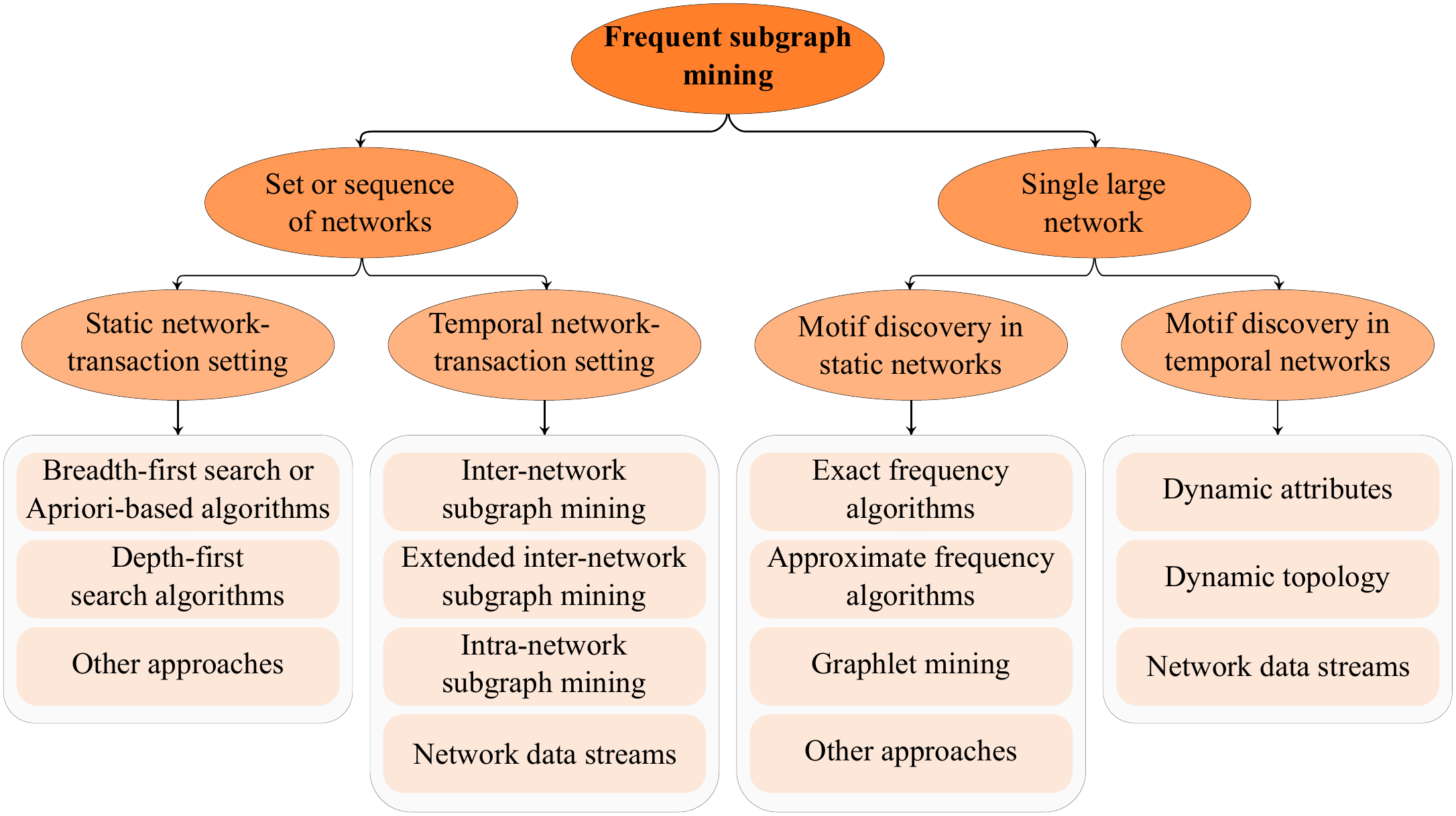}
	\caption{A taxonomy of frequent subgraph mining algorithms proposed in the literature.}
	\label{fig:taxonomy}
\end{figure*}

\subsection{Background}

 Frequent subgraph mining problem has attracted substantial attention in domains where the data can be represented as networks, such as in chemo-informatics \cite{dehaspe1998finding, InokuchiWashioMotoda2000, KuramochiKarypis2001FSG}, health informatics \cite{Ren2019, Sun2019, Cui2018, DU201682, zhang2018subnetwork}, public health \cite{Kavuluru2016, Deng2019, li2017discovering}, bioinformatics  \cite{Mrzic2018bioinformatics, Chen2017, queiroz2020ppigremlin}, social network analysis \cite{Wu2015, Wang2014, Bachi2012}, computer vision \cite{ACOSTA2016, Dammak2014, ACOSTAMENDOZA2012381, BENAOUN2014329, Aoun2011, Aoun2014}, and security \cite{Abusnaina2019, Asrafi2019, Herrera2017, herrera2015framework, Tobias2019}. The frequent subgraph mining in these discplines are either applied to a data set of small networks \cite{jazayeri2021TBD} or a data set of one large network \cite{jazayeri2020motif}. These tasks are traditionally called \textit{network-transaction setting} and \textit{motif discover}y, respectively. Also, the output of mining process is called \textit{frequent subgraphs} (similar to the \textit{frequent itemsets} in \textit{frequent itemset mining} literature) for the former setting and \textit{motifs} after a study by Milo et al. \cite{MiloS2002motifs} for the latter setting.

Figure \ref{fig:taxonomy} provides a taxonomy of the algorithms in the frequent subgraph mining literature. These algorithms can be categorized based on the network data available, either a single large network or a set or sequence of networks. These algorithms are then can be categorized based on the temporality of the data. In cases where the data set is composed of a set of static networks, the algorithms can be classified based on the adopted approach for graph traversal and pattern search strategy. In the temporal network case, the algorithms can be classified based on the patterns being mined; either each network in the sequence is mined (inter-network subgraph mining) or the changes occurring between each pair of consecutive networks in the sequence (intra-network subgraph mining). In some algorithms, the inter-network subgraph mining approach is generalized to multiple sequences (extended inter-network subgraph mining). Besides, in some applications, the networks are added to the sequence in real-time, which creates a separate category of algorithms. For further details and a discussion of algorithms in each subcategory, refer to \cite{jazayeri2021TBD}. The algorithms can be classified based on the adopted approach for frequency computations in single static networks in the motif discovery problem. For motif discovery in a large temporal network, the algorithms can be classified based on the temporal changes occurring in the network data, such as in the network's attributes, network topology, or when the network data is provided in real-time \cite{jazayeri2020motif}.

\subsection{Our contribution and paper organization}
One common approach for mining frequent subgraphs in temporal networks is representing the temporal network as a sequence of static networks. This type of representation of temporal networks has attracted some popularity as it can capture the system's temporal aspects to some extent. However, as will be discussed in the following section, adopting this modeling approach creates a computation-expressiveness trade-off. In other words, increasing the expressiveness of the network representation increases computational costs. For reducing the computational cost, we need to sacrifice some of the system's temporal aspects. Due to this fact, when the duration of interactions between system's entities is not identical for all the interactions, the equal-width temporal aggregation approach might over-represent some of the interactions. 

We propose a continuous-time representation and modeling approach for mining frequent patterns in temporal networks to avoid these limitations. We propose a series of algorithms for mining frequent patterns in a data set of $n$ temporal networks. The performance of the proposed algorithm is evaluated using multiple real-world data sets from different disciplines. Therefore, this paper aims to answer the following questions. Given a data set of $n$ temporal networks and a user-defined frequency threshold of $\epsilon$: How can we provide a lossless representation of temporal networks? Also, how can we mine the complete set of temporal patterns with frequencies more than $\epsilon$ under different definitions of graph and subgraph isomorphisms? To answer these questions, we need to introduce multiple novel concepts and heuristics for mining frequent patterns in temporal networks and provide the definitions for some other available concepts. To allow some structural variation tolerance, we use multiple isomorphism definitions. Finally, the algorithm developed for mining frequent patterns in continuous-time temporal networks, \textit{tempowork}, is explained.

\section{Temporal Network Representation}\label{tempnet_rep}

Networks are considered temporal if their components, vertices and edges, or their associated attributes, change over time. We define temporal networks as follows:

\textbf{Temporal Network:} A temporal network $N$ is defined over a range of $\mathbb{W}$ as an ordered pair, $N=(V,E)$, of two sets, $V=\lbrace v_1,v_2,\ldots, v_n\rbrace$ which is the set of vertices of the network and referred to as $V(N)$, and $E=\lbrace e_1,e_2,\ldots, e_m\rbrace \subseteq V \times V$, which is the set of temporal edges of the network and referred to as $E(N)$. An edge $e_k$ is represented as $e_k=\lbrace v_i,v_j,a_i,l_k,a_j,s_k,\delta_k\rbrace$ where:

\begin{itemize}
\item $v_i$ and $v_j$: identifiers of the edge's two end-points. 
\item $a_i$ and $a_j$: attributes of $v_i$ and $v_j$, respectively. These attributes might be different between the same pair of vertices in various interactions. Also, the same vertex might take different attributes in its interaction with other vertices in overlapping intervals.
\item $l_k$: edge attribute, which might differ between the same pair of vertices in various interactions.
\item $s_k$: the starting point of the interaction window in which $e_k$ is active.
\item $\delta_k$: length of the interaction window in which $e_k$ is active.
\end{itemize}

\begin{figure}[htbp]
\centering
    \includegraphics[width=0.79\linewidth]{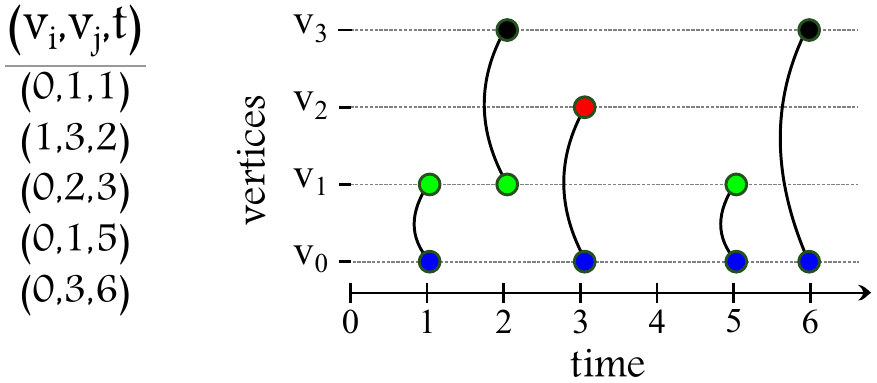}
	\caption{An example of a contact sequence. The list of edges on the left is visualized on the right with an explicit temporal dimension.}
	\label{fig:contact_sequence}
\end{figure}

In some literature, the temporal networks are represented as either a contact sequence or an interval network \cite{Holme2012Temporal, Holme2015colloquium, Pan2011temporal}. The contact sequence representation is composed of edges in the form of $\lbrace v_i,v_j,t \rbrace$, where $v_i$ and $v_j$ are identifiers of the edge's two end-points and $t$ is the point in time that these vertex pairs are connected. Figure \ref{fig:contact_sequence} shows an example of a list of edges in a contact sequence and the corresponding visualization.

\begin{figure}[htbp]
\centering
    \includegraphics[width=0.79\linewidth]{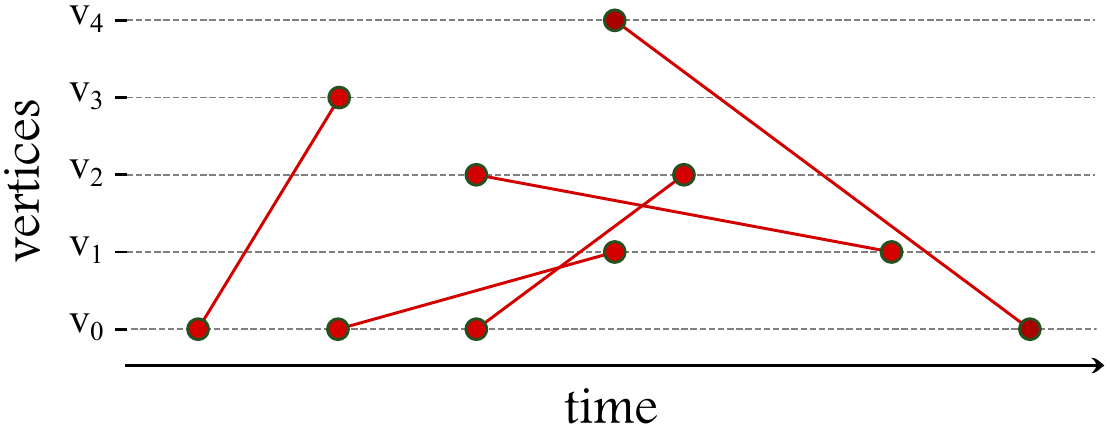}
	\caption{A continuous-time temporal network with vertices and edges active over some period of time.}
	\label{fig:continuous_network}
\end{figure}

In applications where the interactions are instantaneous, adopting a contact sequence representation is preferable (for example, in email correspondence where send and receive events happen in a fraction of seconds). However, in other applications, the duration of interactions is not negligible, for example, in face-to-face interactions, transportation networks, or some of the applications of proximity networks. Therefore, edges are shown as $\lbrace v_i,v_j,s,\delta \rbrace$ (or $\lbrace v_i,v_j,s,f\rbrace$), where $s$ is the start time of the interaction and $\delta$ ($f$) is the duration (finish time) of the interaction. In this case, the contact sequences are a special case in which $\delta=0$ ($s=f$). We adopt this latter representation in this paper and generalize that to both attributed and unattributed networks. Figure \ref{fig:continuous_network} visualizes an unattributed temporal network of this type. This network is composed of five pairs of vertices' interactions over some period of time. Here, for example, we have $v_0$ at the starting time of interaction connected to $v_3$ at the ending time of interaction, meaning that $v_0$ and $v_3$ interact for the entire duration where they are connected.  In Figure \ref{fig:continuous_network}, an undirected network is shown. Therefore, we can change the starting point and ending point to be $v_3$ and $v_0$, respectively. For directed networks, we can start the edges from the tail vertices to the head vertices.

\begin{figure}[htbp]
\centering
\begin{subfigure}[t]{0.8\linewidth}
    \centering
    \includegraphics[width=1\linewidth]{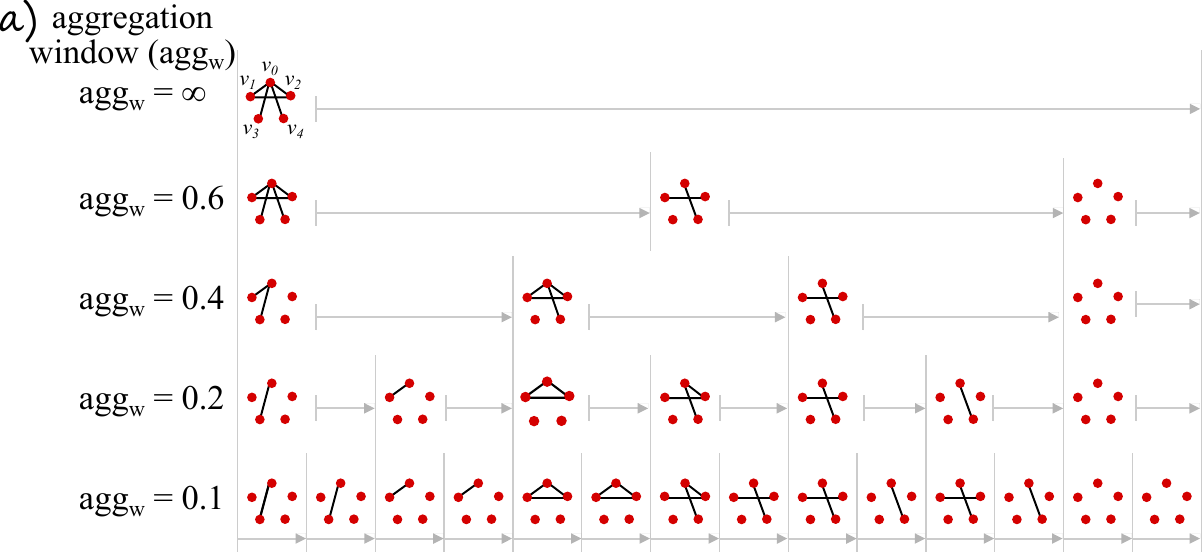}\phantomsubcaption
    \label{fig:temporal_network_seq_a}
\end{subfigure}
\begin{subfigure}[t]{0.8\linewidth}
    \centering
    \includegraphics[width=1.02\linewidth]{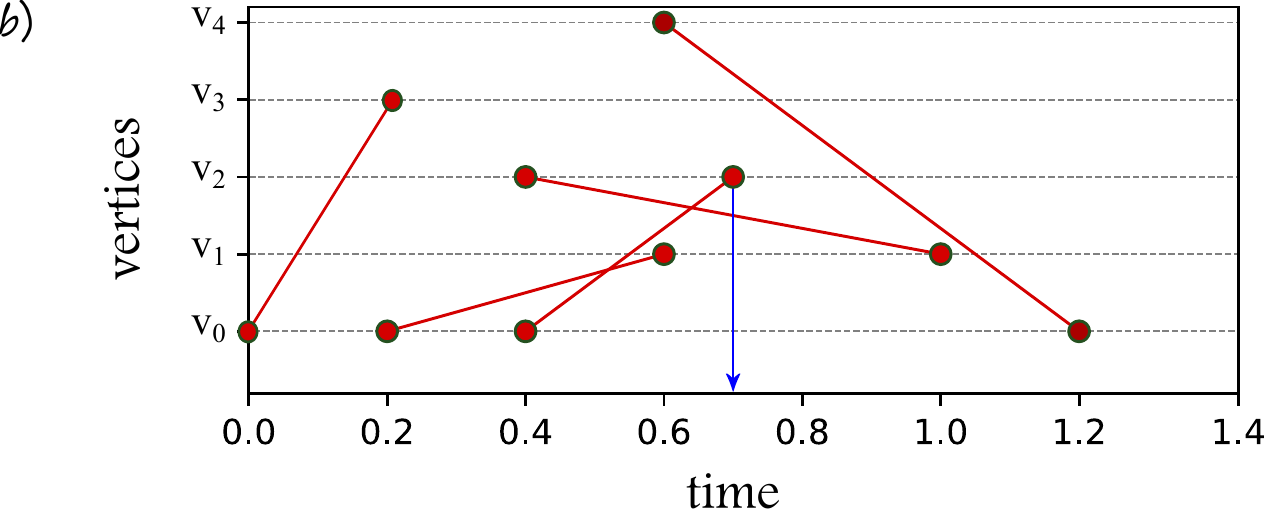}\phantomsubcaption
    \label{fig:temporal_network_seq_b}       
\end{subfigure}
\caption{Temporal network representation in sequential and continuous-time forms. \ref{fig:temporal_network_seq_a}: Sequences of static networks at different aggregation times representing a temporal network. \ref{fig:temporal_network_seq_b}: Continuous-time representation of the same temporal network.}
\label{fig:temporal_network_seq}
\end{figure}

In the literature of frequent subgraph mining, on the other hand, the common approach toward temporal network representation and analysis is converting the temporal dimension to a sequence of intervals and representing the continuous network as a sequence of aggregated static networks \cite{jazayeri2021TBD}. In this representation, for the range of temporal network, $\mathbb{W}$, and aggregation window, $agg_w$, the number of aggregation windows or static networks would be $|seq|=\lceil\mathbb{W}/agg_w\rceil$. For each aggregation window, the relationship between each pair of vertices is aggregated. In other words, for each pair of vertices, a connecting edge is assumed if at least there is one connection between the pair of vertices in that aggregation window, independently from the duration of the connection. It implies that by increasing the $agg_w$, the probability of having a connection between every two vertices in each aggregation window increases. On the other hand, the number of static networks, $|seq|$, decreases. This representation of temporal network is shown in Figure \ref{fig:temporal_network_seq_a} for multiple $agg_w$ as a sequence of static networks related to the temporal network in Figure \ref{fig:temporal_network_seq_b}. 

When the $agg_w=\infty$, we consider an edge between each pair of vertices if there is at least one connection between these two vertices at some point in $\mathbb{W}$. When the $agg_w=\infty$, all the network's dynamic aspect is overlooked, and the continuous network is represented as a single static network. By decreasing the $agg_w$, more continuity characteristics of the network are captured. This is also shown in Figure \ref{fig:temporal_network_seq_a} as we move from $agg_t=\infty$ to $agg_w=0.1$. However, there are some downsides to this type of representation. For example, in Figure \ref{fig:temporal_network_seq_a}, considering $agg_w=0.2$, the representation can capture most of the dynamics associated with the temporal network except for $t \in [0.6,0.8]$. In this period, the corresponding static network shows a connection between $v_0$ and $v_2$ in $[0.6,0.8]$, which is not correct, as this interaction ends at $t=0.7$. Therefore, the $agg_w=0.2$ over-represents the edge between these two vertices from $[0.6,0.7]$ to $[0.6,0.8]$. This over-representation can be modified by reducing the $agg_w$ to $0.1$. In this case, all the temporal network's dynamic characteristics are correctly captured by static networks at regular time-stamps. However, many duplicate static networks are generated at consecutive time-stamps. These duplications negatively impact both the memory requirements and the processing resources needed for frequent pattern mining. Furthermore, considering that the actual duration of some of the interactions might be more than the $agg_w$, some post-processing might be necessary to evaluate the relationships between edges mined in the sequence of static networks and their corresponding interactions in the original network. To overcome these challenges, we adopt an interval network representation (examples are shown in Figures \ref{fig:continuous_network} and \ref{fig:temporal_network_seq_b}). Besides, because there might be multiple edges between each pair of vertices with different attributes, a starting-point sorted edge-based representation is utilized for each pair of vertices in the form of:

\begin{equation}\label{eq:edge}
\begin{aligned}
v_i =\lbrace v_j:[&(a_i^1,e_l^1,a_j^1,s_i^1,\delta_i^1),(a_i^2,e_l^2,a_j^2,s_i^2,\delta_i^2),\\
&\dots,(a_i^k,e_l^k,a_j^k,s_i^k,\delta_i^k)]\rbrace
\end{aligned}
\end{equation}

Then, the network can be written as follows (and potentially as adjacency lists with extra dimensions for labels and interaction windows):

\begin{equation}\label{eq:temp_net}
T = \lbrace v_i=\lbrace v_j:edge\_list_j \rbrace\rbrace
\end{equation}

where $edge\_list_j$ is the list of edges between vertices $v_i$ and $v_j$ in the form of Equation \ref{eq:edge}. 

Note that there might be multiple edges between each pair of vertices appearing at different intervals in $\mathbb{W}$. The vertices might appear or disappear over the continuous dimension. Besides, the attribute of vertices, $a_i$ and $a_j$, and edges, $l_k$, might change in different interactions, even between the same pair of vertices interacting at different intervals in $\mathbb{W}$.

Some other network representations would be special cases of the proposed representation. For example, one can use identical attributes for the network components; then, it is considered an unlabeled or unattributed temporal network. Or, by considering starting point for $e_i$ as constant and $\delta_i=\infty$ for $i\in \{1,\dots,m\}$, the network would represent a static network. 

\section{Preliminary Concepts and Algorithms}
Based on the proposed representation, the next step is defining and introducing the basic concepts of frequent subgraph mining and concepts needed to adopt or develop for mining frequent patterns in temporal networks. The typical approach adopted by different algorithms for mining frequent subgraphs is composed of multiple steps recursively repeated. First, the frequent single vertices or edges in the network database are identified by comparing these simple subgraphs' frequencies with a user-defined threshold. These frequent subgraphs are merged to create a set of candidates to identify frequent larger subgraphs. Then, the frequent subgraphs either are further combined to form the next group of candidates or are grown using frequent vertices and edges. This approach is recursively repeated until no further growth is possible. We adopt a similar approach for frequent pattern mining in temporal networks. However, because we are using a continuous-time representation, after a subsection in which the basic concepts of this problem are described, we define, introduce, and develop multiple available and novel preliminary concepts and algorithms. Then, the algorithm for frequent pattern mining in continuous-time temporal networks is presented.

\subsection{Basic Concepts}
The definitions provided in this subsection are foundational or general concepts in graph theory or frequent subgraph mining literature. We refer to \cite{Bollobas1998, BondyMurty1976, Diestel2005, GrossYellenZhang2013} for foundational concepts. For other more specific concepts, the corresponding references are provided.

\textbf{Graph isomorphism:} Two networks, $N_1=(V_1,E_1)$ and $N_2=(V_2,E_2)$, are isomorphic $(N_1 \cong N_2)$ if there is a bijective function $I:N_1 \rightarrow N_2$ such that:

\begin{equation}
\lbrace v_i,v_j\rbrace \in E_1 \iff \lbrace I(v_i),I(v_j)\rbrace \in E_2
\end{equation}

\begin{figure}[htbp]
\centering
\includegraphics[scale=2]{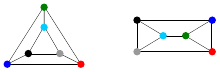}
\caption{Two isomorphic networks. An one-to-one correspondence between the vertices of the two networks are shown with color-coding.}
\label{iso}
\end{figure}

\noindent The function $I$ is called an isomorphism. When the networks are labeled or attributed, the vertices and edges mapped onto each other should have identical labels or attributes. Figure \ref{iso} shows two networks, which, even though they are visualized differently, we can find an isomorphism between them as shown with color-coded vertices. The set of properties identical between isomorphic networks are called graph isomorphism invariants.

\textbf{Subgraph:} Network $N_2=(V_2,E_2)$ is a subgraph of $N_1=(V_1,E_1)$ (and $N_1$ is a supergraph of $N_2$) and shown as $N_2 \subseteq N_1$ if and only if:

\begin{equation}
N_2 \subseteq N_1 \iff V_2 \subseteq V_1 \land E_2 \subseteq E_1
\end{equation}

\textbf{Subgraph isomorphism:} For any given two networks $N_1$ and $N_2$, the problem of finding if $N_1$ includes a subgraph isomorphic to $N_2$ is called subgraph isomorphism problem.\\

\textbf{Subgraph embeddings:} For any given two networks $N_1$ and $N_2$ and $N_2 \subseteq N_1$, $N_1$ may contain one or multiple instances of $N_2$. Any instance of $N_1$ isomorphic to $N_2$ is called an embedding of $N_2$ in $N_1$.

\textbf{Graph and subgraph isomorphism problems' complexities:} Both graph and subgraph isomorphism problems are essential tasks in mining frequent patterns in networks \cite{KuramochiKarypis2001FSG}, and both are computationally expensive. The graph isomorphism problem is not known to be in the P or NP-complete \cite{Fortin96thegraph, Read1977isomorphism}. Although special classes of subgraph isomorphism problems, such as subtree isomorphism, can be solved in polynomial time, in general, the subgraph isomorphism problem is in NP-complete \cite{GareyJohnson1980}. 

Although it is theoretically possible to adopt a brute-force enumeration approach, the subgraph isomorphism problem becomes exponentially expensive as the number of vertices and edges increases. In \cite{Lee2012isomorphism}, some of the subgraph isomorphism algorithms are implemented and compared. They show that different algorithms might perform better in different circumstances. The algorithms proposed in the literature for frequent subgraph mining adopt different sets of heuristics to either avoid this problem or minimize the corresponding computational costs.

\textbf{Canonical labeling:} For a given network, the canonical labeling is defined as a unique code generated independently from and invariant to the representation of the network \cite{Fortin96thegraph}. There are different approaches for generating canonical labeling of networks (For example, refer to \cite{KuramochiKarypis2001FSG, KuramochiKarypis2004FSG, KuramochiKarypis2004FSGtechreport, InokuchiWashioMotoda2000, matsuda2002BGBI, HuanWangPrins2003FFSM}). Once the canonical labelings of networks are generated, they can be used to compare and order networks. The isomorphic networks would have identical canonical labeling. Therefore, as canonical labeling can be used for graph isomorphism verification, the complexity associated with canonical labeling is the same as graph isomorphism. However, adopting different heuristics, the proposed algorithm in the literature using different graph isomorphism invariants to reduce the computational costs of canonical labeling.

\textbf{Pruning mechanisms:} Algorithms proposed in the literature adopt various strategies to narrow down the search space and tackle the complexities discussed above. These strategies are called pruning mechanisms. The most adopted mechanism is the downward closure property (also called antimonotonicity, or  \textit{Apriori} principle \cite{agrawal1994fast}). In the context of network mining, it is defined as if $s_1$ is a supergraph of $s_2$, the frequency of $s_1$, $freq(s_1)$, in a data set should not be more than the frequency of $s_2$:

\begin{equation}
\forall s_2 \subseteq s_1 \implies freq(s_2)\geqslant freq(s_1)
\end{equation}
 
Figure \ref{downward} visualizes an example in which the downward closure property is not met. Various algorithms might adopt different approaches to make this property holds, for example, by defining specific frequency measures, such as the subgraphs embeddings should not have any vertex or any edge in common \cite{VanetikShimonyGudes2006, VanetikGudesShimony2002}. The algorithms adopting these approaches use a frequency-based pruning mechanism. In other words, these algorithms remove non-frequent subgraphs, knowing that none of the candidates created from these non-frequent subgraphs in the subsequent iterations would be frequent. Another pruning mechanism is size-based. Algorithms adopting this mechanism remove candidate subgraphs as soon as their sizes exceed a user-defined size (based on the number of vertices and edges) \cite{BorgeltBerthold2002}. Another important mechanism that can narrow down the search space is structural pruning. In this case, the proposed algorithms remove candidate subgraphs whose isomorphic networks are already evaluated \cite{BorgeltBerthold2002, YanHan2002gSpan}. 

\newcommand{\notimplies}{\;\not\!\!\!\implies}
\begin{figure}[htbp]
\centering
\includegraphics[scale=0.3]{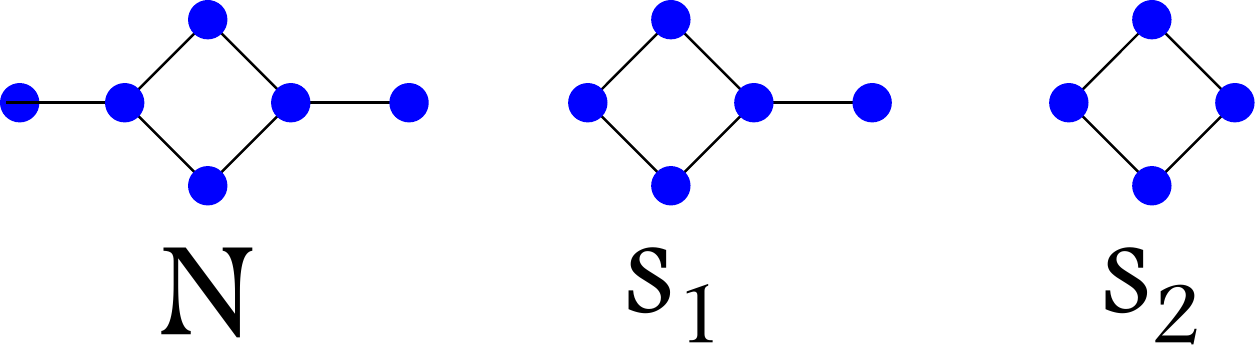}
\caption{Network $N$ and its two subgraphs $(s_2 \subseteq s_1 \subseteq N)$. $freq(s_1)=2$ and $freq(s_2)=1$ in $N$. Here, $(s_2 \subseteq s_1)  \notimplies freq(s_2)\geqslant freq(s_1)$.}
\label{downward}
\end{figure}

\subsection{Constrained Interval Graph}
An essential task in frequent subgraph mining for generating candidates is network traversal. We propose the construction of \textit{Constrained Interval Graph} (CIG), developed on top of the definition of interval graphs for traversing over the temporal networks. An interval $I$ denoted by $[\underline{x},\overline{x}]$ as a subset of the real line is defined as follows:

\begin{equation}
I = [\underline{x},\overline{x}] = \{z \in \mathbbm{R}|\underline{x} \leq z \leq \overline{x}\}
\end{equation}

We consider an interval closed if it includes both end-points of the interval \cite{zwillinger2002crc}. Then, given two intervals, $I=[\underline{x},\overline{x}]$ and $I'=[\underline{y},\overline{y}]$, the intersection of two intervals would be considered empty if $\overline{y} < \underline{x}$ or $\overline{x} < \underline{y}$, otherwise it is defined as follows \cite{moore1966interval}:

\begin{align*}
I \cap I' &=\{z|z\in I \wedge z\in I' \}\\
		  &=[max\{\underline{x},\underline{y}\},
		  min\{\overline{x},\overline{y}\}]
\end{align*}

\textbf{Interval Graph:} Given a set of intervals, $S(I)$, an interval graph $IG$ is defined as a network composed of:
\begin{itemize}
\item $V_{IG}$, in which each vertex of $IG$ is associated with an interval in $S(I)$, and
\item $E_{IG}$, in which an edge represents two connected vertices if the intersection of their corresponding intervals in $S(I)$ is non-empty \cite{fulkerson1965}.
\end{itemize}  

\begin{figure}[htbp]
\centering
\includegraphics[scale=0.9]{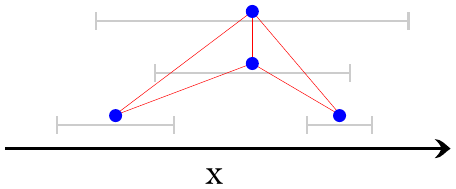}
\caption{An interval graph created from four intervals.}
\label{intervalgraph}
\end{figure}

Figure \ref{intervalgraph} shows an interval graph constructed from four intervals. The vertices of intersecting intervals are connected with an edge.

\begin{figure}[htbp]
\setlength{\lineskip}{0pt}
\centering 
\includegraphics[scale=2.5]{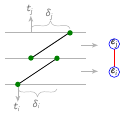} 
\caption{The constrained interval graph created for a temporal network composed of two overlapping edges, $e_i$ and $e_j$ with one vertex in common shown with two blue vertices and a red edge.} 
\label{fig:cig_1} 
\end{figure}

\textbf{Constrained Interval Graph:} For any given edge $e_i$ in a temporal network $N$, the edge and its corresponding end-points are associated with an interval, $\mathcal{I}_i = [s_i,s_i+\delta_i]$. 
We can transform temporal networks into a particular type of interval graph by using the intervals associated with edges and their corresponding end-points. For each edge in the temporal network $N$, we add one node to the interval graph. Then, for each pair of overlapping edges $e_i$ and $e_j$ in $N$, we connect their corresponding nodes in the interval graph if they have end-points representing the same vertex. In other words, using the definition of interval graphs, the constrained interval graph $CIG$ is defined as a network composed of a set of vertices, $V_{CIG}$, which are the edges in the temporal network $N$ and an edge set, $E_{CIG}$, composed of edges connecting pairs of overlapping edges' in the temporal network if they have a vertex in common. Figure \ref{fig:cig_1} shows how a temporal network composed of two edges is transformed to a $CIG$, as the two edges $e_i$ and $e_j$ overlap, and they have one vertex in common.

\begin{figure}[htbp]
\setlength{\lineskip}{0pt}
\centering 
\includegraphics[scale=2.5]{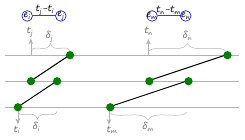} 
\caption{The constrained interval graph created from two temporal networks, each is composed of two overlapping edges with one vertex in common.} 
\label{fig:cig_2} 
\end{figure}

Furthermore, we utilize edges in $E_{CIG}$ to capture the differences between starting points of edges in $N$. For this purpose, an attribute is added to each edge of $E_{CIG}$ computed as the difference between the starting points of the corresponding temporal edges in $N$ connected by an edge in $E_{CIG}$, i.e., $max(s_i,s_j)-min(s_i,s_j)$. Figure \ref{fig:cig_2} shows examples of two temporal networks and their associated $CIG$s composed of one edge with an attribute representing the differences between starting points of the two temporal edges.

\begin{figure}[htbp]
\setlength{\lineskip}{0pt}
\centering 
\includegraphics[scale=2.5]{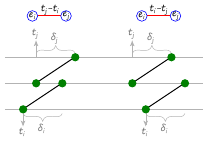} 
\caption{The associated constrained interval graphs are identical for two different temporal networks.} 
\label{fig:cig_symmetry1} 
\end{figure}

Although this definition of constrained interval graph creates a deterministic representation of the corresponding temporal network, it suffers from the edges' delay's symmetric nature. Figure \ref{fig:cig_symmetry1} visualizes this problem. In this figure, two edges $e_i$ and $e_j$ are shown in two configurations wherein the left configuration, $e_i$ appears before $e_j$, and in the right configuration $e_j$ appears first. In both cases, the edge attribute would be identical, as the difference between starting points of the two edges is the same. In other words, given a $CIG$ without the information related to the starting point of edges and vertices, it is impossible to reconstruct the original temporal network from the constrained interval graph because none of the current labels and attributes of $CIG$ expresses which edge appears first.

\begin{figure}[htbp]
\setlength{\lineskip}{0pt}
\centering 
\includegraphics[scale=2.5]{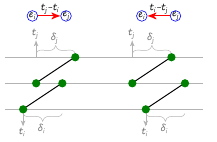} 
\caption{The constrained interval graphs are created with directed edges to differentiate two temporal networks.} 
\label{fig:cig_asymmetry2} 
\end{figure}

To solve this problem, we make the edges in $CIG$ directed, e.g., connecting the vertex related to the edge with the smaller starting point to the vertex pertaining to the edge appearing later in the temporal network by a directed edge. As shown in Figure \ref{fig:cig_asymmetry2}, now the two different temporal networks are uniquely represented by two constrained interval graphs. 

\begin{figure}[htbp]
\setlength{\lineskip}{0pt}
\centering 
\includegraphics[scale=0.5]{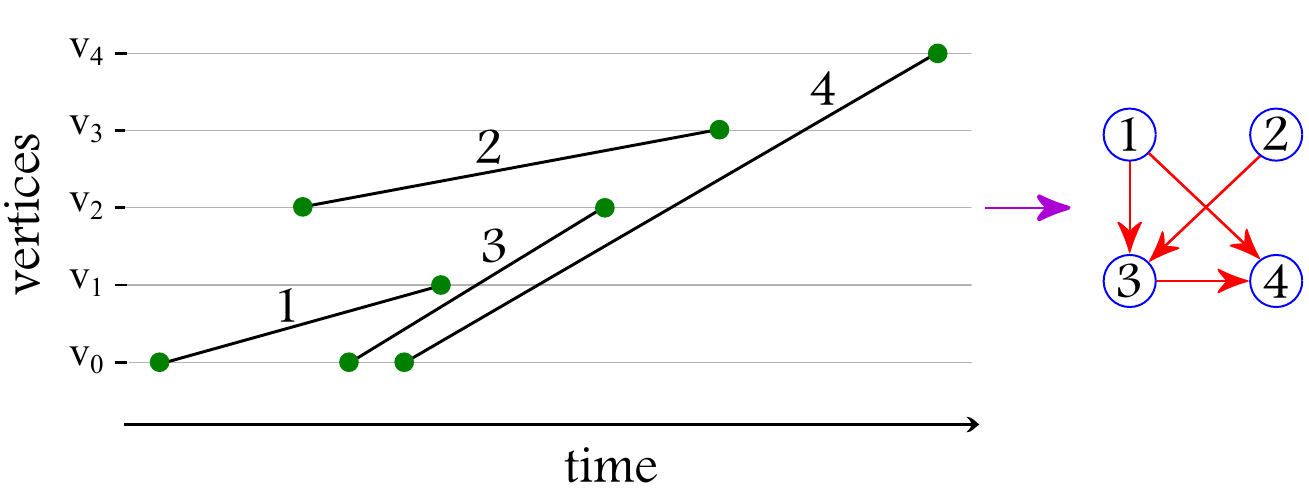} 
\caption{The right network shows the CIG created from the continuous-time temporal network shown on the left. } 
\label{fig:temporal_network_single_CIG} 
\end{figure}

Figure \ref{fig:temporal_network_single_CIG} shows the $CIG$ associated with a continuous-time temporal network. Each vertex of the $CIG$ represents one edge of the temporal network. The direction of edges in $CIG$ shows the relative delay between starting points of each pair of edges in the temporal network. The edges are attributed with these delays. However, for the sake of clarity, these attributes are not shown in this figure.

We assume that the temporal network data available as a list of temporal edges, $\mathcal{L}$, in which the edges are sorted based on their starting time. Therefore, to read the data and construct the corresponding $CIG$, we need to iterate over $\mathcal{L}$ edge by edge. For each temporal edge $e_i$ in $\mathcal{L}$, we need to perform two checks:

\begin{itemize}
\item whether any of the edges appeared earlier in $\mathcal{L}$ has a vertex in common with one of the two end-points of $e_i$, and if yes,
\item whether any of those edges appeared earlier overlap with the interval associated with $e_i$.
\end{itemize}

Considering that there might be multiple other edges passing these checks for each edge, we need to create an efficient data structure to keep track of each vertex and its associated intervals. Once we read a new edge, the data structure should be updated with the new edge information. The data structure adopted for this purpose is the interval tree. Therefore, to efficiently accomplish reading and updating the interval trees, we use a map of vertex \textit{id}s to interval trees, $
\mathcal{M}=\{id:IT_{id}\}$, where the keys of $\mathcal{M}$ represents the identifier of vertices in $N$, and the values of $\mathcal{M}$ are interval trees keeping track of intervals associated with edges having vertex $id$ as one of their end-points. When we read a new edge $e_i$ with two vertices $v_m$ and $v_n$, we update the interval trees mapped to $v_m$ and $v_n$ with the interval of $e_i$, $[s_i,s_i+\delta_i]$. In the following, the characteristics of interval trees are described. Then, we explain the CIG construction step in detail.

\subsection{Interval Tree} For the construction of interval trees for the temporal network's vertices, we follow the approach proposed in \cite{CormenThomasH2009ItAT}. The interval tree data structure can be considered an augmented data structure constructed on top of red-black trees. The red-black trees are (approximately) balanced binary search trees with specific properties. The properties should hold for interval trees to make them efficient data structure for implementing different types of operations, including insertion, deletion, and interval search, on sets of intervals. In our application, we focus on interval insertion and interval search. The interval insertion operation is used when we read a new edge from $\mathcal{L}$, and we want to update the interval trees associated with the two end-points of the edge in $\mathcal{M}$. 

\begin{figure}[htbp]
\centering
\includegraphics[scale=0.9]{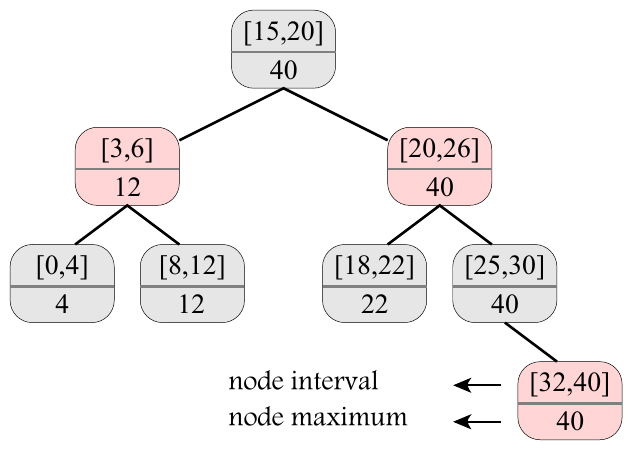}
\caption{An interval tree created from eight intervals. Each node in the interval tree is composed of an interval and maximum value. The node maximum is the maximum values among the right end-point of the node interval, the maximum of the left branch (if any), and the maximum of the right branch (if any). These maximum values are used for interval search and might get updated after each interval insertion. For a detailed discussion of interval insertion and search, refer to \cite{CormenThomasH2009ItAT}.}
\label{intervaltree}
\end{figure}

On the other hand, the interval search is used to connect the end-points of the newly added edge to the end-points of edges already inserted. Therefore, we need to know in which of the earlier appearances of the desired vertices, their corresponding intervals overlap with the new edge's interval. Figure \ref{intervaltree} provides an example for an interval tree created from eight intervals. The interval search and insertion are explained in more detail below.

\textbf{Interval Search} When we read a new edge, $e_i=\lbrace v_m,v_n,a_m,l_i,a_n,s_i,\delta_i\rbrace$, we need to find all the other intervals that overlap with the $e_i$'s interval, $\mathcal{I}=[s_i,s_i+\delta_i]$. This step aims to find all the other vertices that can be connected to the vertex representing $e_i$ in the corresponding $CIG$. Therefore, we need to define a function that searches the $IT_n$ and $IT_m$ separately for all the intervals overlapping with $\mathcal{I}$. If there is no interval in $IT_m$ (or $IT_n$) overlapping with $\mathcal{I}$, then the function should return $NIL$, meaning that either $v_m$ ($v_n$) has not been active in any other edges before $e_i$ or none of the edges having $v_m$ ($v_n$) as one of their vertices overlap with $\mathcal{I}$. Otherwise, the function retrieves a list of edges that have both $v_m$ ($v_n$) as one of their vertices and overlap with $\mathcal{I}$. We are interested in finding all the overlapping intervals with the desired interval when we execute the search operation. If there are $k$ intervals in the interval tree overlapping with the desired interval, the operation can be accomplished in $\mathcal{O}(k \log n)$  \cite{CormenThomasH2009ItAT}. For example, if we want to search for the interval $\mathcal{I}=[9,11]$ in $IT$ provided in Figure \ref{intervaltree}, first, we compare $\mathcal{I}$ with the root of $IT$. Because the root's interval does not overlap with the $\mathcal{I}$, we first check the left branch of $IT$. Because it is not $NIL$ and the node maximum of the left-branch root, $12$, is higher than the left end-point of $\mathcal{I}$, $9$, we consider this left branch $IT_l$ as an interval tree (otherwise, the right branch is considered as the target interval tree), and perform the search on that. We follow this procedure until we either find an interval intersecting with $\mathcal{I}$, or we find that there is no node in the interval tree with an interval intersecting with $\mathcal{I}$. In this example, after choosing the left branch $IT_l$, because the root of $IT_l$ does not overlap with $\mathcal{I}$, we check the left branch of $IT_l$. Because the node maximum of the left branch of $IT_l$, $4$, is less than the left end-point of $\mathcal{I}$, we go to the right branch of $IT_l$ and check the node with the interval $[8,12]$. Because this interval overlaps with $\mathcal{I}$, we return this node. When we want to find all the overlapping intervals with $\mathcal{I}$, we keep searching the right branches as well, even when the conditions related to left branches are met.

\textbf{Interval Insertion}  When we read a temporal network edge by edge, we need to insert the interval $\mathcal{I}=[s_i,s_i+\delta_i]$ associated with the edge $e_i$ in two interval trees $IT_m$ and $IT_n$ associated with the end-point vertices, $v_m$ and $v_n$, of the edge $e_i$. It is proved in \cite{CormenThomasH2009ItAT} that insertion operation can be accomplished in $\mathcal{O}(\log n)$, where $n$ is the number of vertices in the tree. The insertion of a new interval in an interval tree needs more work than an interval search. First, we need to find the correct path from the root to the leaf of an existing interval tree to add the new interval. We color the node of the new interval \textit{red}. However, to keep the data structure efficient, we might need to rotate the interval tree and recolor the nodes. As a result of this operation, the maximum values of nodes might change, as each node maximum, in addition to the node's interval, depends on the node maximum values of the left and right branches (if any) of the node.

\subsection{Constrained Interval Graph Construction}\label{implementation}
Once we have operational functions for insertion and search of intervals in interval trees, we are ready to construct the constrained interval graph, $CIG$, representing the temporal network $N$. For doing that, first, we create an empty map associating vertex identifiers to the interval trees, $\mathcal{M}=\{id:IT_{id}\}$ and initialize $CIG$ with empty sets for vertices and edges. Besides, we assume that the temporal network data is provided as an edge list $\mathcal{L}$ sorted based on the starting points of edges. Then, we iterate over $\mathcal{L}$ edge by edge. For each edge $e_i$, the corresponding interval $\mathcal{I}=[s_i, s_i + \delta_i]$ is created. The edge $e_i$ has two endpoints, $v_m$ and $v_n$. Corresponding to the edge $e_i$, a new vertex, $\mathbbmss{v}_i$, is added to the vertex set of $CIG$. In the next step, we search the $\mathcal{M}$ to see if the interval trees associated with $v_m$ and $v_n$, namely $IT_m$ and $IT_n$, have any intervals overlapping with $\mathcal{I}$. Each of $IT_m$ and $IT_n$ might have none, one, or more than one intervals overlapping with $\mathcal{I}$ representing different edges in $N$. Then, we connect vertices representing these edges of $N$ in $CIG$ to $\mathbbmss{v}_i$ with a directed edge. Each vertex in $CIG$ is labeled with the data of the temporal edge it represents. The labels of vertices in $CIG$ are composed of the labels of the two end-points, $a_i$ and $a_j$, the edge's label, $l_k$, and the length of the interval, $\delta_k$ of the corresponding edge in the temporal network. Each edge in $CIG$ is attributed with the difference in starting points of the associated edges with the vertices (as shown in Figure \ref{fig:cig_asymmetry2}). Next, we update $IT_m$ and $IT_n$ interval trees with the new interval $\mathcal{I}$. After reading all the edges in the edge list, we have the $CIG$ representing the temporal network ready for downstream analysis. Algorithm \ref{cig} provides the pseudo-code for the construction of $CIG$.

\begin{algorithm*}
\caption{CIG Construction Algorithm}\label{cig}
\begin{algorithmic}[1]
\Procedure{Construct\_CIG}{$\mathcal{L}$}
	\State Initialize $CIG=(V,E)$, $V(CIG)=\emptyset$, $E(CIG)=\emptyset$
	\State Initialize $\mathcal{M}=\{id:IT\}$
	\While{$\textrm{!End of }\mathcal{L}$}
		\State Read edge $e_i$ from $\mathcal{L}$ \Comment{$e_i=\lbrace v_m,v_n,a_m,l_i,a_n,t_i,\delta_i\rbrace$}
		\State Define $\mathcal{I} = [t_i,t_i+\delta_i]$
		\State Add one vertex to $CIG$ representing $e_i$ with an associated attribute, ``$a_i\text{-}e_l\text{-}a_j\text{-}\delta_{ij}$''
		\State $v_m\_neighbors \gets IntervalTree\_Search(IT_m,\mathcal{I})$ \Comment{$IT_m = \mathcal{M}[v_m]$}
		\For{$v \in v_m\_neighbors$}
			\State Retrieve interval $\mathcal{I}_v=[t_v,t_v+\delta_v]$ associated with $v$ \hspace*{4em}%
			\State Compute edges' delay $d=t_i-t_v$
			\State Connect vertex including $v$ in $CIG$ to the vertex in $CIG$
			\State \quad representing $e_i$ with a directed edge $e$
			\State Label $e$ with $d$
		\EndFor
		\State $v_n\_neighbors \gets IntervalTree\_Search(IT_n,\mathcal{I})$ \Comment{$IT_n = \mathcal{M}[v_n]$}
		\For{$v \in v_n\_neighbors$}
			\State Retrieve interval $\mathcal{I}_v=[t_v,t_v+\delta_v]$ associated with $v$ 
			\State Compute edges' delay $d=t_i-t_v$ \rlap{\smash{$\left.\begin{array}{p{4cm}c@{}}\\{}\\{}\\{}\\{}\\{}\end{array}\color{red}\right\}\color{red}\begin{tabular}{l}This loop is similar to\\the previous For Loop,\\ with $m \rightarrow n $\end{tabular}$}}
			\State Connect vertex including $v$ in $CIG$ to the vertex in $CIG$ 
			\State  \quad representing $e_i$ with a directed edge $e$
			\State Label $e$ with $d$
		\EndFor		
		\State Call $IntervalTree\_Insert(IT_m,\mathcal{I})$
		\State Call $IntervalTree\_Insert(IT_n,\mathcal{I})$		
	\EndWhile		
	\State return $CIG$ \& $\mathcal{M}$
\EndProcedure	
\end{algorithmic}
\end{algorithm*}

\subsection{Temporal Network Reconstruction}\label{term}
In some of the applications, we need to convert subgraphs of a $CIG$ to the corresponding temporal networks that they represent. It can be easily shown that the relationships between $CIG$s and temporal networks are not one-to-one. In other words, although Algorithm \ref{cig} always constructs a unique $CIG$ for any temporal network given, there might be multiple subgraphs of $CIG$ representing the same temporal network. The mapping of $CIG$ to the corresponding temporal network is accomplished using the attributes of vertices and edges and edge directions. The vertices' attributes in $CIG$ provide the attributes of edges in the associated temporal network. The edges' attributes and directions in $CIG$ are to infer the magnitudes of overlaps and delays between pairs of edges in the temporal network. 

Taking a more in-depth look into the constrained interval graph and the temporal relationships of edges represented by directed edges in this graph, it can be deduced that the constrained interval graph is a directed acyclic graph (as we assume that time has a given direction). 

For reconstructing a temporal network from a $CIG$, we always start with the vertices having the smallest identifiers and proceed in the vertex set. It is because the edge list $\mathcal{L}$ that the $CIG$ is created from in the first place has been sorted based on the starting points of the edges. Therefore the vertices with the smallest identifiers represent the edges that appear earlier in $\mathcal{L}$ and have a smaller starting time. Besides, we assume that the first vertex in $CIG$ has a starting point of zero (or any other arbitrary value). The starting points of other edges are derived from relative delays to the other edges in $CIG$.

The reconstruction of a temporal network from a $CIG$ or a subgraph of $CIG$ is performed as follows. We start with the vertex with the smallest identifier in the (subgraph of) $CIG$, $\mathbbmss{v}_1$. Then we traverse the $CIG$ with one of the directed acyclic graph traversal strategies (such as breadth-first search or depth-first search). We consider the starting point of the edge represented by $\mathbbmss{v}_1$ as zero. Therefore, we create the first edge of the temporal network, $e_1=\lbrace v_1,v_2,a_1,l_1,a_2,0,\delta_1\rbrace$.  Using the attributes of the edges originating from $\mathbbmss{v}_1$, we can find the starting time of the neighbor vertices (representing edges in the temporal network). We traverse the $CIG$ vertex by vertex to generate the temporal network's edges using this strategy. Algorithm \ref{dfs_recon} provides the pseudo-code for reconstructing temporal networks from their associated $CIG$s.

\begin{algorithm*}
\caption{Temporal Network Reconstruction Algorithm}\label{dfs_recon}
\begin{algorithmic}[1]
\Procedure{Reconstruct\_TemporalNetwork}{$CIG$, $support$} \Comment{$support$ is one of $DS$ networks supporting $CIG$}
	\State Initialize $\mathcal{L}$ as an empty list \Comment{This list is used for recording edges of temporal network}
	\ForAll{$v$ in $V(CIG)$}
		\State $v.visited = False$
	\EndFor	
	\State Using the first node read from $CIG$, extract $\{a_1,e_l,a_2,\delta_{12}\}$
	\State Using the $support$, identify the two nodes representing $v_1$ and $v_2$ \Comment{Note: each $v \in CIG$ represent two vertices $v_i$ and $v_j$ of the temporal network, and their connecting edge}
	\State Define $temp\_edge=\{v_1,v_2,a_1,e_l,a_2,0,\delta_{12}\}$
	\State $\mathcal{L}.push\_back(temp\_edge)$
	\State Create an empty stack, $S$
	\State $S.push(v_1)$ 
	\State Create an empty map for recording start times of edges, $STs=\{\}$
	\State $STs=\{v_1:0\}$ \Comment{We consider the starting time of $v_1$ as zero}
	\While{!$S.empty()$}
		\State $v = S.pop()$
		\If{$!v.visited$}
			\State $v.visited = True$
			\ForAll{$u$ in $v.neighbors()$}
				\If{$!u.visited$} \Comment{\textcolor{red}{\refi{\ident{}}{\scriptsize }{\ident{}} Extra code relative to DFS starts here!}}
					\State Extract temporal edge attributes from $u$, $\{a_i,e_l,a_j,\delta_{ij}\}$ 
                	\State Using the $support$, identify the two nodes representing $v_i$ and $v_j$
                	\State $STs[u]=STs[v]+e_{uv}^l$ \Comment{$e_{uv}^l$ is the label of the edge connecting vertices $u$ and $v$ in $CIG$}
                	\State Define $temp\_edge=\{v_i,v_j,a_i,e_l,a_j,STs[u],\delta_{ij}\}$
                	\State $\mathcal{L}.push\_back(temp\_edge)$ \Comment{\textcolor{red}{\refi{\ident{}}{\scriptsize }{\ident{}} Extra code relative to DFS ends here!}}
					\State $S.push(u)$
				\EndIf 
				
			\EndFor
		\EndIf
	\EndWhile
	\State return $\mathcal{L}$
\EndProcedure	
\end{algorithmic}
\end{algorithm*}

\section{The \textit{tempowork} Algorithm}\label{olgu}
After being able to construct constrained interval graphs from temporal networks and reconstruct temporal networks from (any subgraphs of) constrained interval graphs, we can discuss the \textit{tempowork} algorithm for the identification of frequent patterns in temporal networks. Given a data set of temporal networks, $DS$, we convert the $N_i \in DS$ for $i \in \{1,n\}$ to their associated $CIG_i$ and create a secondary data set of constrained interval graphs, $DS'$. The $DS'$ can be considered as a data set of static networks. However, $CIG$s has multiple characteristics which should be noted in the mining process.

Furthermore, we need to adopt a series of strategies for candidate generation, subgraph enumeration, graph isomorphism, and subgraph isomorphism checks and verification. It would be beneficial to adopt strategies such that no duplicates subgraphs are generated, and any brute-force implementation of graph and subgraph isomorphism problems are avoided. 

In the process of frequent pattern mining in a network-transaction setting, the graph isomorphism is performed to verify if any of the candidates generated during the frequent subgraph mining process are identical. The algorithm should either avoid generating duplicate candidates or identify identical candidates as soon as possible in the mining process. On the other hand, the subgraph isomorphism is used when we want to verify if any of the networks in the data set contain the candidates being enumerated. In the worst-case scenario, we need to iterate over the networks in the data set one by one and verify if any of them contain the candidate of interest. Both of these tasks are computationally very expensive. We adopt a set of heuristics to avoid both of them as much as we can.

We start the mining process with the identification of frequent vertices in $DS'$. These vertices represent frequent temporal edges in the $DS$. Besides, we identify frequent edges in $DS'$. Each frequent edge in $DS'$ is composed of two temporal edges representing patterns in networks of $DS$ consisting of two edges with identical delays. The frequent edges can be reordered based on their frequencies. We use these frequent edges (as seeds and in combination with other frequent patterns already detected) to generate new candidates in each iteration of the mining process.

As we discussed, the $CIG$s of temporal networks are directed acyclic graphs, $DAG$s. However, in addition to the characteristics that $CIG$s inherit from $DAG$s, $CIG$s have an essential feature; for mining the $CIG$s, we do not need to mine all the edges. All the frequent subgraphs of temporal networks can be represented with subtrees of the corresponding $CIG$s. This characteristic results from the temporal relationships between vertices in $CIG$ (representing temporal edges in temporal networks). In other words, if some of the temporal relationships between vertices of $CIG$ are known, we might be able to deduce the temporal relationships between other vertices of $CIG$.

Therefore, we start mining by adding one edge at a time to known frequent subgraphs. The first set of frequent subgraphs are frequent edges. We add an edge to a frequent subgraph to create a new candidate; we make sure that this new edge does not create any cycle in the candidate's undirected version. Also, to minimize the number of duplicates generated, we will be using the lexicographic ordering introduced in gSpan \cite{YanHan2002gSpan, YanHan2002gSpantech}. For the sake of space, we do not repeat the principles of and pattern growth strategies based on the lexicographic ordering, as they are perfectly discussed in gSpan papers. However, there are some changes we need to apply to these principles based on the characteristics of DAGs and CIGs that will be addressed in the following.

The CIGs are directed networks. Therefore, we need to consider the direction of edges when we are generating new candidates. We only create new candidates by adding forward-edges, as we are interested in mining non-cyclic patterns. The edges that their addition to the candidates create cyclic subgraphs in undirected versions of CIGs result in duplicate candidates. As discussed, each edge $e$ of the CIGs carries the delay information between temporal edges that the two end-points of the $e$ represent. If these two end-points are already connected through a different path, the delay that $e$ represents can be deduced from the path that connects these two end-points. In other words, we do not need to create candidates using backward edges. This approach eliminates a large number of duplicate candidates. Although adopting the lexicographic ordering and forward growth of candidates prevents the generation of many duplicate candidates, the graph isomorphism problem can not be avoided entirely. In fact, it is even worse than the case where we mine a data set of static networks. It is because two completely non-cyclic candidates might still represent the same temporal network. To make sure that these candidates are detected as early as possible, we first convert the candidates to their temporal networks (using Algorithm \ref{dfs_recon}) and re-create their CIGs (using Algorithm \ref{cig}). If candidates represent identical temporal networks, their corresponding CIGs should be isomorphic. Therefore, by recording the canonical labeling of the CIGs, we can make sure no duplicate candidates are generated and used for pattern growth. The canonical labeling can be created using the same lexicographic ordering principles or any other known to be efficient approaches proposed in the literature, such as nauty and Traces \cite{MCKAY201494}, or bliss \cite{bliss2007}.

We create an embedding list for each frequent subgraph for frequency computation. When we add a forward edge to a frequent parent subgraph to create a new child candidate, we only check the embeddings that support the parent subgraph (instead of searching all the networks in the data set). Suppose the new edge added to a parent subgraph creates a valid child subgraph. In that case, we check whether the embeddings supporting the parent subgraphs can be extended in their associated networks with the new edge. If the list of networks that support the child subgraph meets the user-defined support threshold, we consider the child frequent, which can be used for candidate generation in the next iterations. Otherwise, neither the child nor the candidates created from this child (based on the downward-closure property) would be frequent and can be eliminated. We can avoid the subgraph isomorphism problem altogether by adopting this strategy for subgraph enumeration and frequency computation.  Algorithm \ref{tw} provides the pseudo-code for the \textit{tempowork} algorithm.

\begin{figure*}[htbp]
\centering
    \includegraphics[width=1\linewidth]{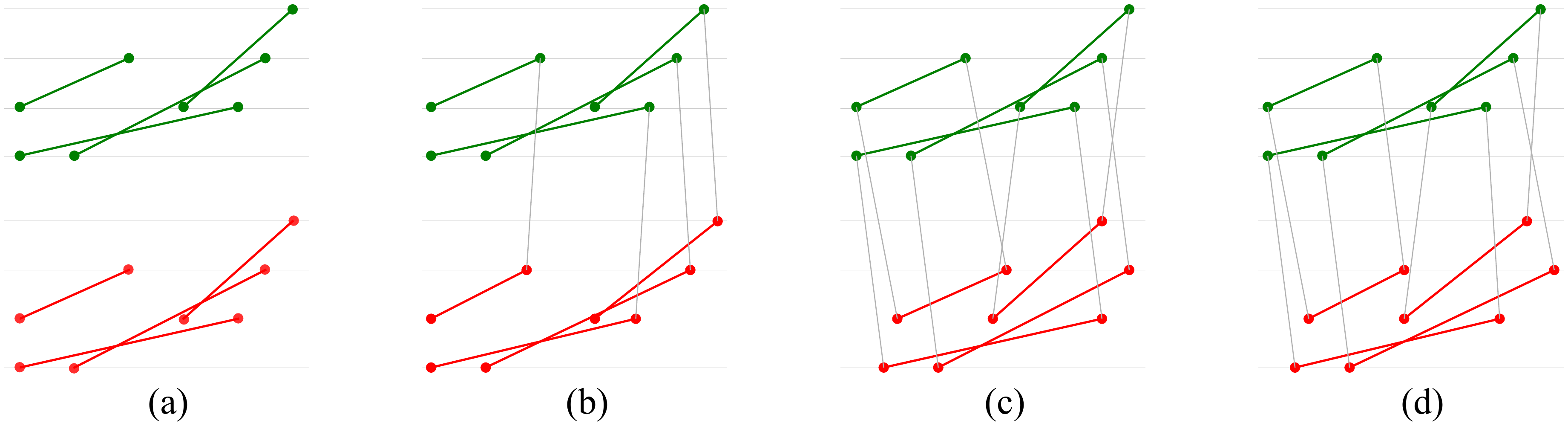}
	\caption{A schematic representation of the four different isomorphism definitions considered for the temporal networks. In the exact formulation (a), all the temporal edges' durations and the inter-edge delays are identical between the two networks. In the inexact-time version (b), some user-defined variations in edge durations are permitted. However, the magnitudes of delays between edge pairs in the two isomorphic networks are exactly identical. In the exact-time sequence-preserved formulation (c), the edges mapped to one another have exactly identical durations. However, two networks are considered isomorphic if the edges' appearances follow the same sequences in both networks. In the inexact-time sequence-preserved formulation (d), the constraint of duration equality of corresponding edges is removed based on a user-defined variation tolerance mechanism.}
	\label{fig:isomorphisms}
\end{figure*}

The temporal information of the temporal networks is recorded in the CIGs. The duration of the temporal edges ($\delta$s) are recorded in the vertices of CIGs, and the delays between temporal edges ($\Delta$s) are recorded in the edges of CIGs. Both of these temporal characteristics might be noisy in real-world applications. Therefore, we can relax the temporal constraints of isomorphisms at different levels to allow some noise tolerance when we create the embedding lists. In the following, we define four types of graph isomorphism based on the level of tolerance we consider two networks isomorphic. A visual representation of these definitions is provided in Figure \ref{fig:isomorphisms}.

\textbf{Exact-time graph isomorphism:} Two networks, $N_1=(V_1,E_1)$ and $N_2=(V_2,E_2)$, are exact-time isomorphic $(N_1 \overset{e}{\simeq} N_2)$ if there is a bijective function $I^e$ such that in addition to the constrained defined for graph isomorphism, for any $\lbrace v_i,v_j\rbrace \in E_1$ mapped to $\lbrace I^e(v_i),I^e(v_j)\rbrace \in E_2$ and any pair of edges $e_i,e_j \in E_1$ mapped to $I^e(e_i), I^e(e_j) \in E_2$:

\begin{equation}
\begin{aligned}
\delta \lbrace v_i,v_j\rbrace = \delta \lbrace I^e(v_i),I^e(v_j)\rbrace\\
\Delta(e_i,e_j) = \Delta(I^e(e_i),I^e(e_j))
\end{aligned}
\end{equation}

Where $\delta \lbrace v,v'\rbrace$ represents the duration of the edge connecting $v$ and $v'$ and $\Delta(e,e')$ denotes the delay between starting points of any pair of edges $e,e'$.

\textbf{Inexact-time graph isomorphism:} Two networks, $N_1=(V_1,E_1)$ and $N_2=(V_2,E_2)$, are inexact-time isomorphic $(N_1 \overset{i}{\simeq} N_2)$ if there is a bijective function $I^i$ such that in addition to the constrained defined for graph isomorphism, for any $\lbrace v_i,v_j\rbrace \in E_1$ mapped to $\lbrace I^i(v_i),I^i(v_j)\rbrace \in E_2$ and any pair of edges $e_i,e_j \in E_1$ mapped to $I^i(e_i), I^i(e_j) \in E_2$:

\begin{equation}
\begin{aligned}
\delta \lbrace v_i,v_j\rbrace \simeq \delta \lbrace I^i(v_i),I^i(v_j)\rbrace\\
\Delta(e_i,e_j) = \Delta(I^i(e_i),I^i(e_j))
\end{aligned}
\end{equation}

\textbf{Exact-time sequence-preserved graph isomorphism:} Two networks, $N_1=(V_1,E_1)$ and $N_2=(V_2,E_2)$, are exact-time sequence-preserved isomorphic $(N_1 \overset{es}{\simeq} N_2)$ if there is a bijective function $I^{es}$ such that in addition to the constrained defined for graph isomorphism, for any $\lbrace v_i,v_j\rbrace \in E_1$ mapped to $\lbrace I^{es}(v_i),I^{es}(v_j)\rbrace \in E_2$ and any pair of edges $e_i,e_j,e_k \in E_1$ mapped to $I^{es}(e_i), I^{es}(e_j), I^{es}(e_k) \in E_2$:

\begin{equation}
\begin{aligned}
&\delta \lbrace v_i,v_j\rbrace = \delta \lbrace I^{es}(v_i),I^{es}(v_j)\rbrace\\
&\Delta(e_i,e_j) < \Delta(e_j,e_k) \iff  \\
&\Delta(I^{es}(v_i),I^{es}(v_j)) < \Delta(I^{es}(v_j),I^{es}(v_k)) \\
&\Diamond(\Delta(e_i,e_j) \neq \Delta(I^{es}(v_i),I^{es}(v_j))) \wedge \\
&\Diamond(\Delta(e_j,e_k) \neq \Delta(I^{es}(v_j),I^{es}(v_k))
\end{aligned}
\end{equation}

Here, the expressions starting with $\Diamond$ means that the magnitude of delays can differ between corresponding isomorphisms. The sequence of appearance of edges is identical among the two networks.

\textbf{Inexact-time sequence-preserved graph isomorphism:} Two networks, $N_1=(V_1,E_1)$ and $N_2=(V_2,E_2)$, are inexact-time sequence-preserved isomorphic $(N_1 \overset{is}{\simeq} N_2)$ if there is a bijective function $I^{is}$ such that in addition to the constrained defined for graph isomorphism, for any $\lbrace v_i,v_j\rbrace \in E_1$ mapped to $\lbrace I^{is}(v_i),I^{is}(v_j)\rbrace \in E_2$ and any pair of edges $e_i,e_j,e_k \in E_1$ mapped to $I^{is}(e_i), I^{is}(e_j), I^{is}(e_k) \in E_2$:

\begin{equation}
\begin{aligned}
&\delta \lbrace v_i,v_j\rbrace \simeq \delta \lbrace I^{is}(v_i),I^{is}(v_j)\rbrace\\
&\Delta(e_i,e_j) < \Delta(e_j,e_k) \iff  \\
&\Delta(I^{is}(v_i),I^{is}(v_j)) < \Delta(I^{is}(v_j),I^{is}(v_k)) \\
&\Diamond(\Delta(e_i,e_j) \neq \Delta(I^{is}(v_i),I^{is}(v_j))) \wedge \\
&\Diamond(\Delta(e_j,e_k) \neq \Delta(I^{is}(v_j),I^{is}(v_k))
\end{aligned}
\end{equation}

In these definitions, we need to provide user-defined thresholds to show how much noise is tolerable for identifying inexact time equivalences. Also, we can discretize the duration of edges. The discretization problem has its own rich literature \cite{discretization2016}. Different characteristics of the networks (such as attribute of vertices and edges and classes of networks) can be used to perform discretization in supervised or unsupervised modes. To preserve the sequences, we can assume that all the edges in the CIGs have the same attribute and let the differences between the duration of edges and their starting points create the sequences of edge appearances.

\begin{algorithm*}
\caption{\textit{tempowork} Algorithm}\label{tw}
\begin{algorithmic}[1]
\Procedure{Preprocessing}{$DS$, $min\_supp$, $ISO\_type$}
	\State Initialize $DS'$  \Comment{This is the CIG data set}
	\State Initialize $1\_None\_map$ \Comment{This is a dictionary: \{node\_labels: location of their embeddings in $DS'$\}}
	\State Initialize $1\_Edge\_map$ \Comment{This is a dictionary: \{edge\_labels: location of their embeddings in $DS'$\}}
	\ForAll{$T$ in $DS$}
		\State $CIG_T=Construct\_CIG(T)$ \Comment{If $ISO\_type \in \{i,es,is\}$, relabel $\delta$s and/or $\Delta$s accordingly}
		\State $DS'.push(CIG_T)$
	\EndFor	
	\State Order node labels (``$a_i\text{-}e_l\text{-}a_j\text{-}\delta_{ij}$''s) and edge labels ($\Delta$s) descendingly based on their frequencies in CIGs of $DS'$
	\State Remove node and edge labels in CIGs with frequencies $< min\_supp$
	\State Relabel lexicographically node and edge labels in CIGs with frequencies $\geq min\_supp$
	\State Update $1\_None\_map$ and $1\_Edge\_map$ dictionaries with the location of their embeddings in $DS'$
	\State Return $1\_None\_map$, $1\_Edge\_map$
\EndProcedure	
\Procedure{\textit{tempowork}}{$DS$, $min\_supp$, $ISO\_type$}
	\State Initialize $frequent\_patterns$ \Comment{This is a dictionary: \{frequent\_patterns: location of their embeddings in $DS'$\}}
	\State Initialize $cl\_list$ \Comment{It is a list for recording list of canonical labeling of $CIG$ already mined}
	\State $1\_None\_map$, $1\_Edge\_map = Preprocessing(DS, min\_supp, ISO\_type)$
	\State Update $frequent\_patterns$ with $1\_None\_map$ and $1\_Edge\_map$
	\State $temp\_1\_Edge\_map = 1\_Edge\_map$ \Comment{We create a copy of the $1\_Edge\_map$}
	\Procedure{CIG\_Mining}{$pattern$, $pattern\_support$}	
		\ForAll{$edge$ in $temp\_1\_Edge\_map$}
			\State $new\_pattern = \text{add }edge \text{ to } pattern \text{ as forward edge}$
			\If{$new\_pattern$ has the minimum lexicographical form}
				\State $temp\_tw = Reconstruct\_TemporalNetwork(new\_pattern, support)$ \Comment{$support$ is one of $DS$ networks supporting $new\_pattern$ created from the function input argument $pattern\_support$}
				\State $temp\_cig = Construct\_CIG(temp\_tw)$
				\State $cl = canonical\_labeling(temp\_cig)$ \Comment{This line computes the canonical labeling of $temp\_cig$}
				\If{$cl$ not already in $cl\_list$}
					\State $cl\_list.push(cl)$
					\State $frequent\_patterns[new\_pattern] = new\_pattern\_support$
					\State Call CIG\_Mining($new\_pattern$, $frequent\_patterns[new\_pattern]$)
				\EndIf
			\EndIf
			
		\EndFor
	\EndProcedure
	\ForAll{$edge, frequent\_patterns[edge]$ in $1\_Edge\_map$}
		\State Call CIG\_Mining($edge$, $frequent\_patterns[edge]$)
		\State Remove $edge$ from $temp\_1\_Edge\_map$
	\EndFor	
	\State Return $frequent\_patterns$
\EndProcedure	
\end{algorithmic}
\end{algorithm*}

\section{Experiments}
To the best of our knowledge, this is the first algorithm proposed for mining frequent patterns in continuous-time temporal networks. We evaluated the proposed algorithm's performance using three real-world data sets with different numbers of vertices, edges, and time windows. Furthermore, we used different definitions of isomorphism and evaluated their impacts over the number of frequent patterns identified and computation time. We also used different discretization parameters to evaluate the impact of noise tolerance on the number of frequent subgraphs detected by the algorithm. In the following, the data sets are described. Table \ref{dataset_characteristics} summarizes these three data sets' characteristics. Then, the results obtained by applying the proposed algorithm to these data sets are provided, and the findings are discussed.

\begin{table}[htbp]
\caption{Characteristics of the data sets used for the experiments, including the number of networks in each data set and the average number of vertices $\overline{|V|}$ and edges $\overline{|E|}$ of networks in continuous-time representations of the data sets.}
\label{dataset_characteristics}
\centering
\small
\begin{tabular}{p{4.8cm}p{0.6cm}p{1.2cm}p{0.7cm}}
\hline
data set & $|N|$ & $\overline{|V|}$ & $\overline{|E|}$\\
\hline
Hospital ward proximity network & 5 & 48 & 2806\\
High school contact networks & 7 & 151 & 2824\\
Sepsis EHR data set & 13,229 & 5 & 24 \\
\hline
\end{tabular}
\end{table}

\subsection{Data Sets}
\subsubsection{Hospital Ward Proximity Network} 
The first data set is a proximity network created using wearable RFID devices from contacts between patients and healthcare providers in a hospital ward in Lyon, France \cite{Vanhems2013, SocioPatterns}. Two vertices are considered proximate if they are in a range of 1–1.5 meters from one another. The data collection has been performed over five days, December 6, 2010 (1:00 pm)-December 10, 2010 (2:00 pm). During these five days, the contacts are recorded at 20-second intervals. In total, the network's vertices represent 75 individuals (paramedical staff: 27; patients: 29; medical doctor: 11; administrative staff: 8). However, only a subset of these vertices is active in each of these five days (Table \ref{dataset_characteristics}).

\subsubsection{High school Contact Networks} 
This data set's networks represent seven days of temporal interactions among high school students of five classes in Lycée Thiers, Marseilles, France \cite{High_School2014, SocioPatterns}. The interactions are recorded, adopting a proximity face-to-face encounters approach by using wearable sensors. The networks' vertices represent students, and there is an edge between each pair of students if their sensors exchange radio packets in 20-second intervals. We consider each day as a separate network. The objective is to identify the frequent interaction patterns among students on different days. 

\subsubsection{Sepsis EHR Data Set} 
This data set is composed of retrospectively collected EHR data related to cellular and physiological responses of sepsis patients in 13,229 visits. Our previous work \cite{Jazayeri2019} showed that patterns of simultaneous failures of cellular and physiological responses and biomarkers are different among subpopulations of sepsis (sepsis with and without septic shock). Furthermore, our findings showed that considering these co-failures can predict patients' mortality more accurately than individual failures. Therefore, in this experiment, we are interested in identifying temporal patterns of simultaneous failures of responses and biomarkers occurring more frequently in patients with sepsis. Further description of this data set and criteria used to identify sepsis patients are provided in Appendix A.

\subsection{Implementation} 
\subsubsection{Hospital Ward Proximity Network} 
\begin{figure}[htbp]
\centering
\includegraphics[width=0.7\linewidth]{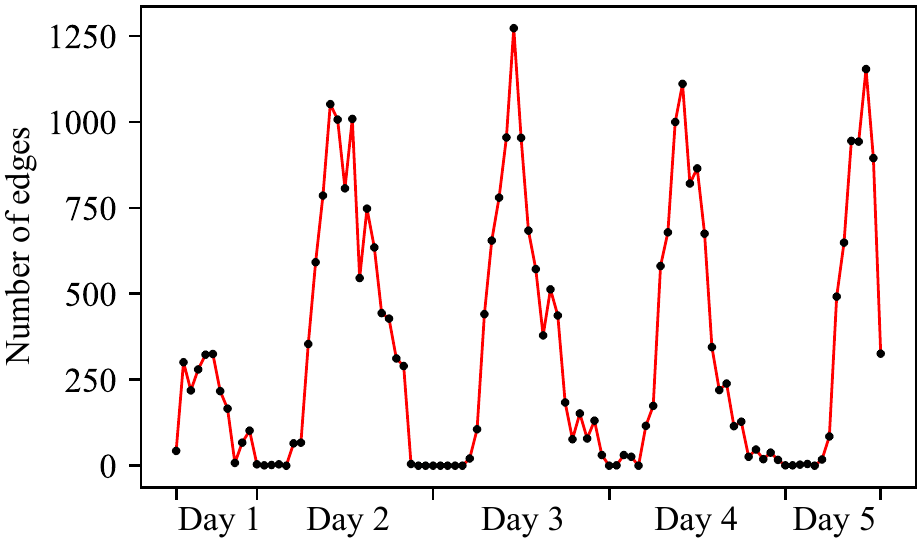}
\caption{Number of edges in the hospital ward proximity network over five days at 20-second intervals.}
\label{ward_network_5days}
\end{figure}

\begin{figure}[htbp]
\centering
\includegraphics[width=0.7\linewidth]{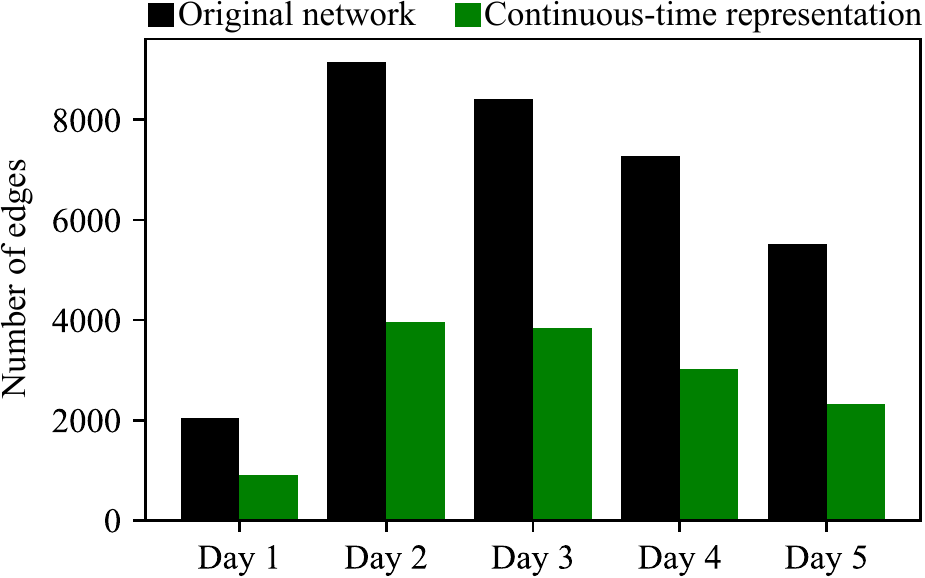}
\caption{Number of edges in the original data set at 20-second intervals and in the continuous-time representation proposed in this paper.}
\label{continuous_representation}
\end{figure}

The visualization of the number of edges (in 20-second intervals) during each hour of these five days showed that the number of interactions has its maximum in noon and minimum at around midnight (Figure \ref{ward_network_5days}). Therefore, we created one temporal network for each day (For the first and last day, the data collection is performed for 11 and 14 hours, respectively). Besides, we pre-processed the data to create continuous-time temporal networks. Because in continuous-time temporal network representation, all the interactions between the same pair of vertices in consecutive intervals are reported by one edge with an interaction interval, this pre-processing reduced the number of edges (Figure \ref{continuous_representation}). Also, the vertices are labeled based on their associated groups (paramedical staff, patients, medical doctors, and administrative staff).

\subsubsection{High school Contact Networks}  
The interactions between students in this data set are recorded with a 20-second resolution at the most detailed temporal level. However, the interactions might last more than 20 seconds. Therefore, we pre-processed the data into the continuous-time temporal network representation proposed in this paper. For this data set, we label vertices in two different ways. First, the vertices are labeled with anonymous student identifiers (ids). Therefore, the networks' vertices are uniquely labeled. In the second scenario, the class labels are used as labels for their corresponding students. Therefore, students of each of the five classes have identical labels. These two scenarios help evaluate and compare the nature of frequent patterns and computation time of the algorithm for unlabeled and labeled networks.

\subsubsection{Sepsis EHR Data Set} 
Each network in this data set represents one visit of a patient with sepsis. The maximum number of nodes in each visit is 19, representing different cellular and physiological responses or biomarkers. The responses and biomarkers are used to identify sepsis occurrences. The edges of the network represent the co-failures of these responses or biomarkers. In this data set, the vertex identifiers are used as vertex labels; therefore, all the vertices are uniquely labeled.

The proposed algorithm was applied to the three network data sets described above using multiple frequency thresholds. Also, we implemented the algorithm for different definitions of isomorphism with different discretization parameters. The results are shown in Figures \ref{results_all}, \ref{discret}, and \ref{patterns}. The numerical values of results are provided in Appendix B-Table 2. The findings are discussed in the following subsection. The Python implementation of the algorithm used in this study is publicly available from the PyPI repository and can be installed under ``\textit{tempowork}''. The experiments were conducted on a personal computer with an Intel Core i7-8700 3.20GHz CPU processor with 16.0 GB installed RAM.

\begin{figure*}[htbp]
\centering
\includegraphics[width=0.95\linewidth]{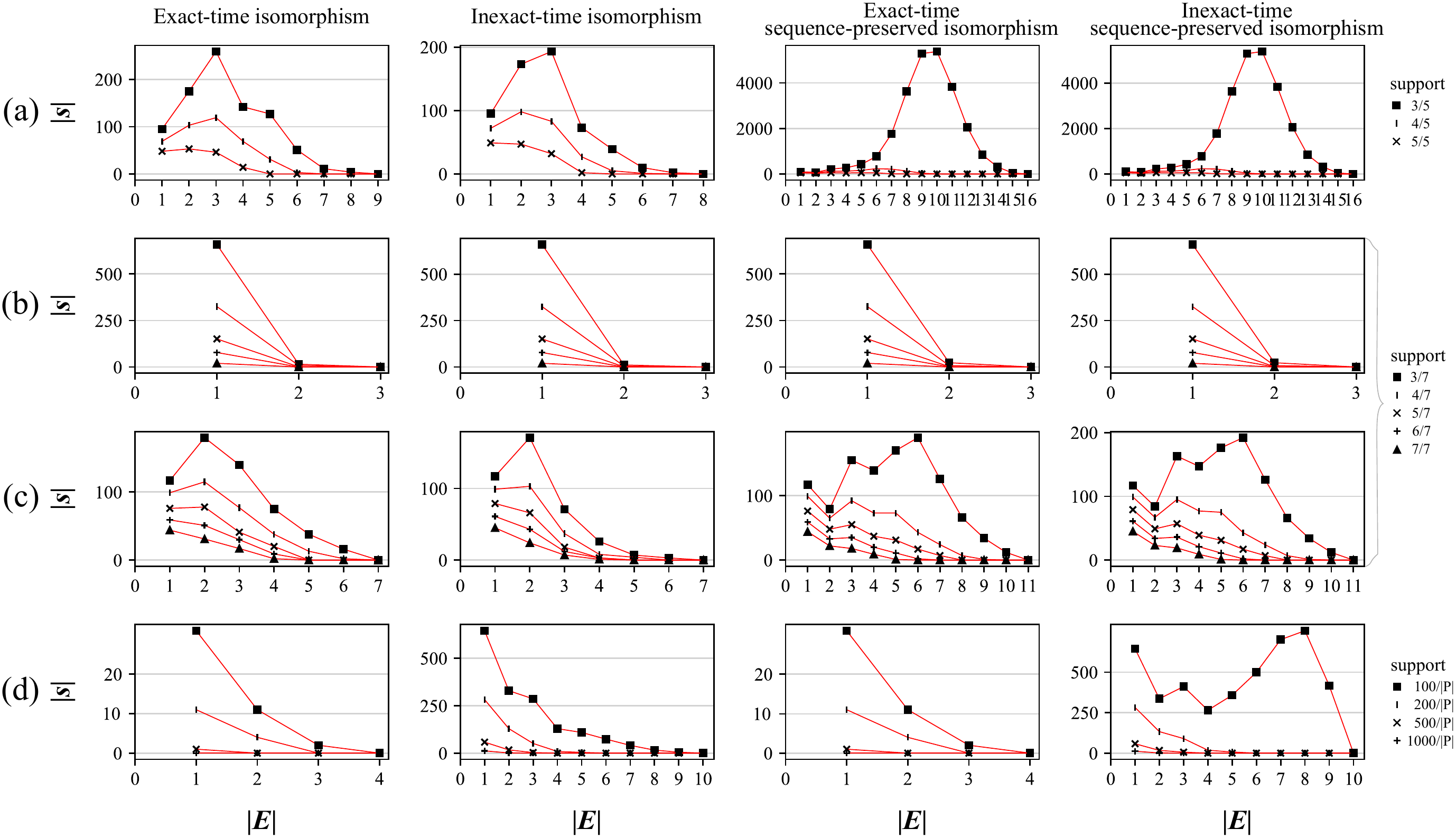}
\caption{The number of frequent patterns detected by the algorithm using different definitions of isomorphism from (a) hospital ward proximity, (b) high school contact networks with ids as labels, (c) high school contact networks with classes as labels, and (d) sepsis EHR data sets at different support thresholds ($|E|$: number of edges in the frequent subgraphs, $|s|$: number of frequent subgraphs detected, $|p|$: total number of patients in the sepsis EHR data set, 13,229).} 
\label{results_all}
\end{figure*}

\begin{figure*}[htbp]
\centering
\includegraphics[width=0.95\linewidth]{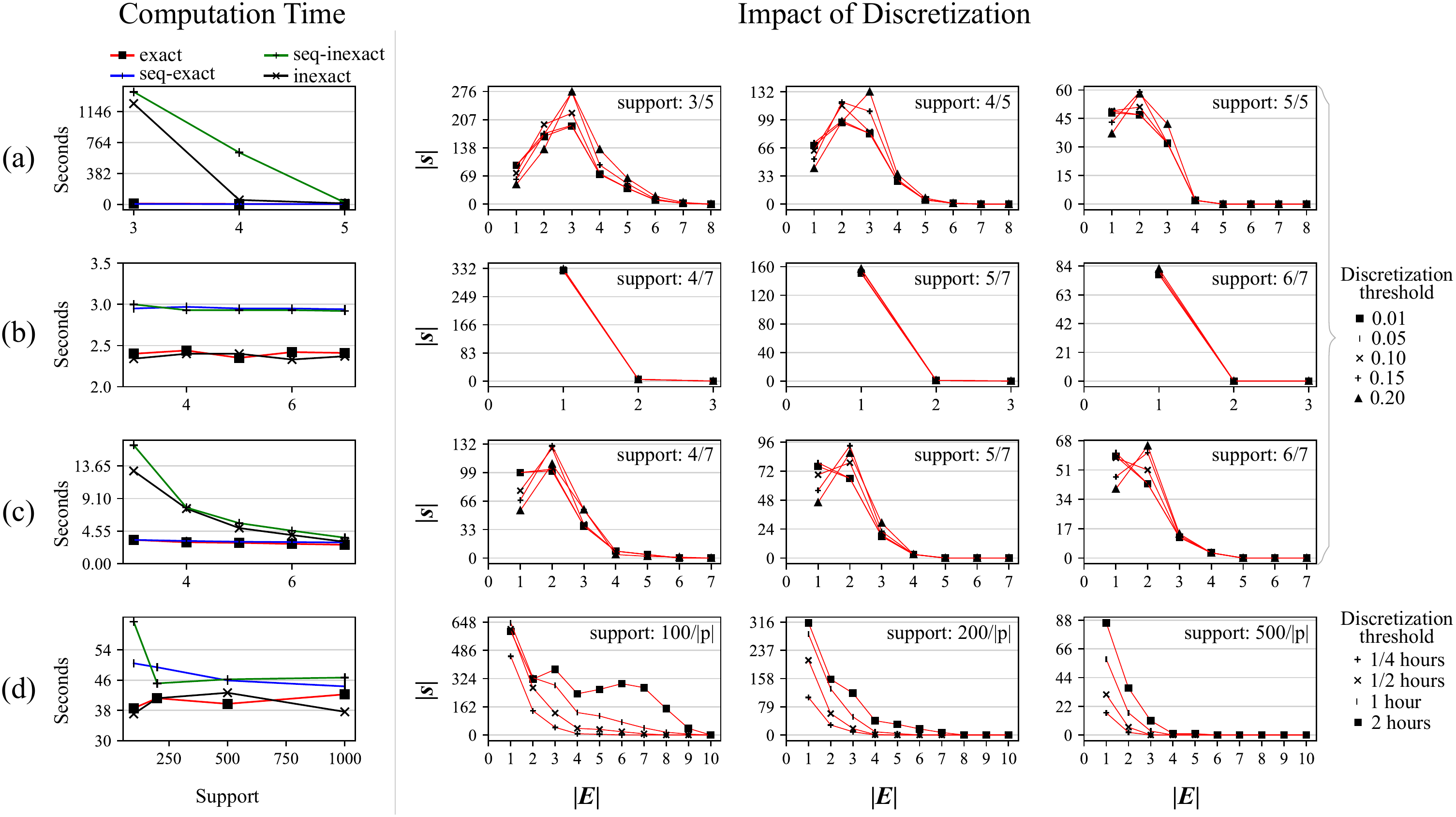}
\caption{The computation time (left) and impact of different discretization parameters (right) on the number of frequent subgraphs detected at different support thresholds for (a) hospital ward proximity networks, (b) high school contact networks with ids as labels, (c) high school contact networks with classes as labels, and (d) sepsis EHR data sets. For discretization parameters, the inexact version of isomorphism is used.  ($|E|$: number of edges in the frequent subgraphs, $|s|$: number of frequent subgraphs detected, $|p|$: total number of patients in the sepsis EHR data set, 13,229).}
\label{discret}
\end{figure*}

\begin{figure*}[htbp]
\centering
\includegraphics[width=\linewidth]{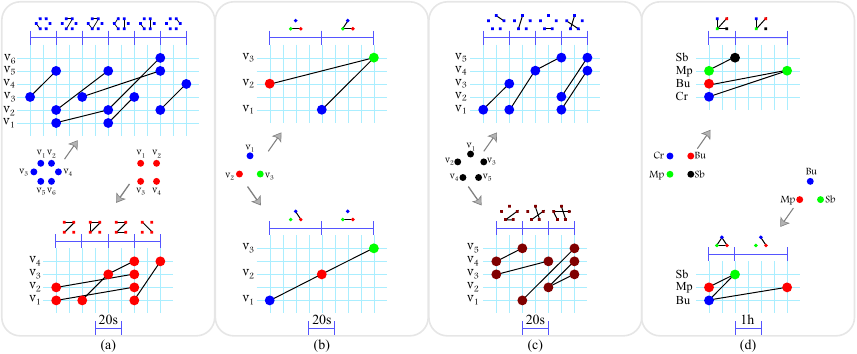}
\caption{Some of the frequent patterns detected in (a) hospital ward proximity networks, (b) high school contact networks with ids as labels, (c) high school contact networks with classes as labels, and (d) sepsis EHR data sets. Vertices with the same labels are shown with identical colors. Each row represents one vertex. The interactions of pairs of vertices over time are shown as edges spanning from left to right. Also, for each of the continuous-time temporal patterns detected, a typical network representation is provided above the pattern. Note that when these patterns are frequent, any subgraph of these patterns is frequent as well, based on the downward closure property (s: seconds, h: hour, Bu: Blood urea nitrogen, Cr: Creatinine, Mp: Mean arterial pressure, Sb: Systolic blood pressure).}
\label{patterns}
\end{figure*}

\subsection{Discussion}
The proposed algorithm was applied to the pre-processed data sets. We used the four different types of isomorphisms, different support values, and different discretization parameters. The results show that increasing the support thresholds decreases the number of frequent patterns in all three data sets consistently (Figures \ref{results_all}). 

The results of the implementation for the high school contact networks when the ids are used as labels showed that most of the frequent patterns are composed of either one temporal edge (the interactions between two specific students for a particular period) or two temporal edges (the interactions among three specific students in different arrangements) (Figures \ref{results_all}-(b)). This can be attributed to the unique labels considered in this implementation for each student. Also, the algorithm did not find many frequent subgraphs for the sepsis EHR data set when it is implemented in the exact-time or exact-time sequence-preserved modes (Figures \ref{results_all}-(d)). It can be attributed to the nature of this data set. The lab measurements are recorded at times with a precision of fractional portions of a second. At this precision, it is very rare to find patterns with identical durations among patients. However, when we implement the algorithm for the high school contact networks with classes as labels (Figures \ref{results_all}-(c)) or for sepsis EHR data set for other two modes of isomorphism (Figures \ref{results_all}-(d)), the algorithm could find many frequent patterns at different support thresholds.

It should be noted that although adoption of inexact-time isomorphism increased the number of frequent patterns detected, it might decrease the total number of frequent patterns in some cases. In fact, inexact-time isomorphism has two opposing effects. It may group some infrequent patterns into one frequent pattern (if they are isomorphic based on the inexact-time isomorphism definition). In this case, the number of frequent subgraphs increases. Simultaneously, some of the different frequent subgraphs detected based on the exact-time isomorphism might be considered one subgraph based on the inexact-time isomorphism definition, consequently decreasing the total number of frequent subgraphs. 

In all the data sets, we observed that using the sequence-preserved definitions almost always results in a higher number of frequent patterns compared to their exact-time and inexact-time counterparts, respectively. It is because we are relaxing the constraints related to the overlaps between edges, as far as they follow the same sequence of edge appearances.

Implementing the algorithm for different discretization parameters shows that these variations impact data sets differently (Figure \ref{discret}). In the first and second data sets, using different discretization parameters does not significantly change the number of frequent patterns. However, due to the nature of the sepsis EHR data set and the temporal precision in data collection, we do not have many frequent exactly identical patterns common among patients. However, as we increase the discretization parameters and allow higher tolerance of noises, we observe many frequent patterns among the patients. Also, it should be noted that the opposing effect of inexact-time isomorphism discussed earlier can affect the number of frequent subgraphs detected for various discretization parameters differently as well.

The three data sets used in this study have some significant differences that directly impact the frequent subgraph mining computation time. The first data set is composed of five networks. The average number of vertices and edges in this data set is 48 and 2,806, respectively, and each vertex is labeled with one of the four possible labels. On the other hand, although the second data set comprises seven networks (about the same number of networks as the first data set), it is composed of a larger number of vertices (about three times more than the first data set) with almost the same number of edges (and nearly the same number of vertex labels). This difference between the number of vertices in these two data sets results in the lower density in the second data set and consequently decreases the computation time (Figure \ref{discret}-(a) vs. Figure \ref{discret}-(c)). When the vertices in the second data set are labeled with unique ids of students, it results in a lower number of frequent patterns and, accordingly,  significantly lower computation time (Figure \ref{discret}-(b) vs. Figure \ref{discret}-(c)). In the sepsis EHR data set, we have a maximum of 19 vertices in each of 13,229 networks. However, all the vertices are uniquely labeled with one of the associated cellular and physiological responses or biomarkers they represent. Also, not all of them are necessarily present in the patients' hospitalization records. The labeling approach significantly reduces the computation time required for performing the graph isomorphism (Figure \ref{discret}-(d)). 

Figure \ref{patterns} shows two sample frequent subgraphs detected in the four implementations of the algorithm on the three data sets. Nearly all the frequent subgraphs detected in the first data set were composed of paramedical staff interacting with each other (Figure \ref{patterns}-(a)-top). However, the algorithm also detected a few frequent patterns among medical doctors \ref{patterns}-(a)-bottom). The exact-time instances of these patterns are observed in at least three (out of 5) networks of the data set. Figure \ref{patterns}-(b) and (c) show four frequent temporal patterns observed in at least 4 (out of 7) networks of the second data set in the ids as labels and class as labels implementations, respectively. The algorithm detected a few frequent patterns in the second data set with ids as labels with more than one temporal edge. These patterns are among three students interacting in different formations, such as three specific students interact for some particular time at the same time, one student joins the other two students somewhere in between of the first two students' interaction \ref{patterns}-(b)-top), or two students are interacting for some time, and one of them ends this interaction and start another interaction with a different student for a while  \ref{patterns}-(b)-bottom. When the students are labeled with classes, there are many frequent patterns with a larger number of temporal edges. However, most of the patterns are among students of the same class  \ref{patterns}-(c). 

\begin{figure}[htbp]
\centering
\includegraphics[width=0.7\linewidth]{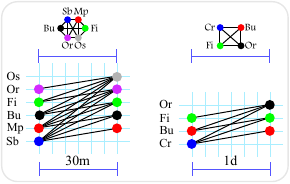}
\caption{Co-failure of multiple cellular and physiological responses in patients with sepsis over the same period of time (m: minutes, d: day, Bu: Blood urea nitrogen, Cr: Creatinine, Fi: Fraction of inspired oxygen ($FiO_2$), Or: $SpO_2/FiO_2$, Os: Oxygen ($O_2$) source, Mp: Mean arterial pressure, Sb: Systolic blood pressure).}
\label{sepsis_patterns}
\end{figure}

Finally, Figure \ref{patterns}-(d) shows two patterns identified as frequent patterns in at least 100 sepsis patients by running the algorithm in the inexact-time isomorphism mode. Although these patterns show some changes in the patterns' edges over time, most of the frequent patterns detected in the sepsis EHR data set are related to the simultaneous failures of multiple cellular and physiological responses over some period of time (Figure \ref{sepsis_patterns}). It can be attributed to the nature of laboratory test measurements performed almost simultaneously in pre-defined intervals. These patterns can also be considered complete networks, as all the vertices are connected with one another. Both patterns are identified using the algorithm in the inexact-time sequence-preserved setting, one with 15 minutes duration deviation tolerance (left) and the other one with one-hour duration deviation tolerance (right).  

\section{Conclusion}
This study proposed a novel approach for mining the complete set of frequent patterns in continuous-time temporal networks. A novel representation of temporal networks for frequent pattern mining is described. We developed an original method for continuous-time temporal network traversal and considered four types of isomorphism to detect frequent subgraphs in a temporal network data set. Applying the proposed algorithms on three real-world data sets showed that the proposed algorithm could identify patterns that cannot be detected using previously proposed algorithms.

One avenue for future research would be mining frequent patterns in network data streams. Most of the previous work defines network streams as updating batches of network components at discrete intervals. Investigating the frequent pattern mining problem, including the data structure requirements, for temporal networks with continuous updates of network components would be a challenging future direction with a wide variety of applications. 

In our future work, we want to use the outcome of the proposed algorithm for mining the evolution among frequent patterns, where patterns might emerge, merge, shrink, or grow. Another applied work that we want to explore is investigating the significance of frequent patterns detected in predicting the outcome of patients with different health conditions and the progression of their complications.


%

%

\ifCLASSOPTIONcompsoc
  \section*{Acknowledgments}
\else
  \section*{Acknowledgment}
\fi

This work is supported in part by the National Science Foundation under the Grant NSF-1741306, IIS-1650531, and DIBBs-1443019.  Any opinions, findings, and conclusions or recommendations expressed in this material are those of the author(s) and do not necessarily reflect the views of the National Science Foundation.

\ifCLASSOPTIONcaptionsoff
  \newpage
\fi



%

\bibliographystyle{IEEEtran}
\bibliography{references}

%

\begin{IEEEbiography}[{\includegraphics[width=1in, height=1.25in, clip, keepaspectratio]{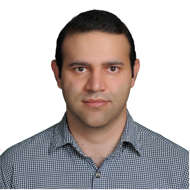}}]{Ali Jazayeri} is a Ph.D. candidate in the College of Computing and Informatics (CCI) at Drexel University, and his dissertation focuses on mining frequent substructures and their evolution in temporal networks. Ali is an instructor in the MS in Data Science program at CCI. He is working in the Healthcare Informatics Research Laboratory under the supervision of Professor Christopher C. Yang. 

He received his BS in  Materials Science \& Engineering from the Amirkabir University of Technology in 2006 and his MS in Socio-Economic Systems Engineering and Sociology from Sharif University of Technology and the University of Tehran in 2010 and 2016, respectively.  He has over 20 publications in IEEE Transactions on Big Data, Journal of Complex Networks, IEEE International Conference on Healthcare Informatics, ACM Conference on Bioinformatics, Computational Biology, and Health Informatics, and more. 
\end{IEEEbiography}

\begin{IEEEbiography}[{\includegraphics[width=1in, height=1.25in, clip, keepaspectratio]{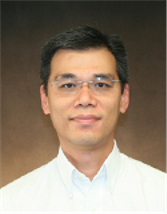}}]{Christopher C. Yang} is a professor in the College of Computing and Informatics at Drexel University. He also has a courtesy appointment at the School of Biomedical Engineering, Science, and Health Systems. He is the Director of Data Science Programs and the Program Director of MS in Health Informatics. His research interest includes data science, artificial intelligence, machine learning, healthcare informatics, social media analytics, electronic commerce, and intelligence and security informatics. He has over 340 publications in top-tier journals, conferences, and books, such as ACM Transactions on Intelligent Systems and Technology, ACM Transaction on Management Information Systems, IEEE Transactions on Knowledge and Data Engineering, IEEE Transactions on Computational Social Systems, PLOS One, Journal of Medical Internet Research, Artificial Intelligence in Medicine, and more. 

His work has been supported by NSF, NIH, PCORI, HK RGC, etc. He is the editor-in-chief of Journal of Healthcare Informatics Research and Electronic Commerce Research and Application. He is the editor of the CRC book series on Healthcare Informatics and the founding steering committee chair of the IEEE International Conference on Healthcare Informatics. He has been the general chair of over 5 conferences and program chairs of over 10 conferences.

He is the director of the Healthcare Informatics Research Lab. His recent research includes pharmacovigilance, drug repositioning, predictive modeling of sepsis, comorbidity and co-medication analysis,  predictive modeling of disengagement driving for injury prevention, heterogeneous network mining, distributed graph computing, health intervention through social media for substance use disorders, and social network analytics.
\end{IEEEbiography}




\end{document}


\pagestyle{plain}
\appendix
\section{Sepsis EHR Data Set}
\subsection{Study Population}
This dataset is derived from a study composed of 210,289 visits related to 119,968 patients admitted to two hospitals with 1,100 total in-hospital beds of a single tertiary care health care system between July 2013 and December 2015. Other inclusion criteria are age $\geq$18 at arrival and with visit types of inpatient, Emergency Department only (outpatient), or observational visits. The study was approved by the health systems Institutional Review Board.

\subsection{Sepsis Patients Identification}
From the study population, we identified infected patients who have experienced sepsis. The infection criteria are defined as being administered with anti-infective for at least four days or a positive viral PCR (polymerase chain reaction) test for influenza. The infected patients who have experienced organ dysfunction in the period of 24 hours before the first anti-infective administration until the last administration are considered sepsis patients. There are seven organ systems considered. These organ systems, their associated cellular and physiological responses, and the criteria used for organ dysfunctions identification are shown in Table 1. Besides, vasopressor administration was considered a sign of organ dysfunction as well. We also considered three biomarkers, Procalcitonin, C-Reactive Protein, and Erythrocyte Sedimentation Rate, to identify sepsis patients. This set of criteria were developed by referring to established Sepsis-3 guidelines \cite{Singer2016} and subject matter experts' input. In this study, death was considered as in-hospital death or discharge to hospice care. Using organ dysfunction criteria and death definition, the patients with sepsis were identified and categorized into two groups of survivors and non-survivors. Furthermore, if any of the patients were discharged to hospice or died while receiving an anti-infective in a period less than four days of anti-infective administration, the patient was considered a sepsis non-survivor \cite{Rhee2017}. 

\begin{center}
\begin{table}[htbp]\label{org_dysfun}
\small
\caption{Cellular and physiological responses and biomarkers recorded as individual features, and the criteria resulting in the corresponding organ dysfunction}
\begin{tabular}{p{2.3cm}p{4.3cm}cp{5.3cm}}
\hline\hline
Organ system & Response & Abbreviation & Failure criteria \\
\hline
\multirow{3}{*}{Cardiovascular} & Systolic blood pressure (SBP) & Sb & \textless 90 mmHg \\
 & SBP$_{max}$* - Systolic BP & Sd & \textgreater 40 mmHg within an 8-hour period \\
 & Mean arterial pressure (MAP) & Mp & \textless 65 mmHg \\\hline
\multirow{3}{*}{Renal} & Creatinine & Cr & \textgreater 1.2 mg/dL \\
 & (Creatinine - C$_{base}$**)/(C$_{base}$) & Cd & \textgreater 50\% from initial creatinine \\
 & Blood Urea Nitrogen (BUN) & Bu & \textgreater 20 mg/dL \\\hline
\multirow{2}{*}{Hematopoietic} & WBC & Wb & \textless 4,000 cells/mL \\
 & Platelet & Pl & \textless 100,000 cells/mL \\\hline
Metabolic & Lactate & La & \textgreater 2.0 mmol/L \\\hline
Gastrointestinal & Bilirubin & Bi & \textgreater 2 mg/dL \\\hline
\multirow{4}{*}{Respiratory} & Fraction of inspired oxygen ($FiO_2$) & Fi & \textgreater 21\% \\
 & Pulse oximetry ($SpO_2$) & Px & \textless 90\% \\
 & $SpO_2/FiO_2$ & Or & \textless 421 \\
 & Oxygen ($O_2$) Source & Os & Mechanical ventilation required (bilevel positive airway pressure (BiPAP) or continuous positive airway pressure (CPAP) or ventilator) \\\hline
\multirow{2}{*}{Central Nervous} & Glasgow Comma Score & Gc & \textless 14 \\
 & Glasgow Best Verbal Response & Gv & \textless 5 \\\hline
\multirow{3}{*}{Biomarkers} & Procalcitonin & pc & \textgreater 0.15 ng/mL \\
 & C-Reactive Protein & cr & \textgreater 8 mg/L \\
 & Erythrocyte Sedimentation Rate & sr & \textgreater 20 mm/hr 
\\\hline\hline
\end{tabular}
\par 
\begin{itemize}
 \item[*:] Maximum systolic blood pressure for each observation within 8-hour windows.
\item[**:] Initial creatinine value observed in each visit.
\end{itemize}
\end{table}
\end{center}

\pagebreak
\begin{landscape}
\section{Numerical Results}

%
\begin{scriptsize}
\begin{longtable}{|c|c|c|c|c|c|c|c|c|c|c|c|c|c|c|c|c|c|c|c|} 
\caption{The numerical results of experiments for the three data sets. Iso.: Isomorphism type, Freq.: Frequency threshold, Disc.: Discretization parameter, CT: Computation time (seconds), DS: Data set, s: seconds, m: minutes, h: hours, W: Hospital ward proximity network, H\_1: High school contact networks with ids as labels, H\_2: High school contact networks with classes as labels, S: Sepsis EHR data set. The first 15 columns show the size of subgraphs detected in the corresponding experiment. }\\
\hline
\textbf{1} & \textbf{2} & \textbf{3} & \textbf{4} & \textbf{5} & \textbf{6} & \textbf{7} & \textbf{8} & \textbf{9} & \textbf{10} & \textbf{11} & \textbf{12} & \textbf{13} & \textbf{14} & \textbf{15} & \textbf{Iso.} & \textbf{Freq.} & \textbf{Disc.} & \textbf{CT} & \textbf{DS} \\ 
\hline
\endhead 
\hline
95 & 175 & 259 & 142 & 127 & 51 & 11 & 4 & 0 & 0 & 0 & 0 & 0 & 0 & 0 & e & 3 & 0 & 10.26 & Hospital \\ 
\hline
69 & 103 & 119 & 69 & 31 & 3 & 0 & 0 & 0 & 0 & 0 & 0 & 0 & 0 & 0 & e & 4 & 0 & 5.76 & Hospital \\ 
\hline
48 & 53 & 46 & 14 & 0 & 0 & 0 & 0 & 0 & 0 & 0 & 0 & 0 & 0 & 0 & e & 5 & 0 & 3.9 & Hospital \\ 
\hline
95 & 166 & 191 & 73 & 39 & 10 & 2 & 0 & 0 & 0 & 0 & 0 & 0 & 0 & 0 & i & 3 & 0.01 & 5.25 & Hospital \\ 
\hline
95 & 173 & 193 & 73 & 39 & 10 & 2 & 0 & 0 & 0 & 0 & 0 & 0 & 0 & 0 & i & 3 & 0.05 & 5.41 & Hospital \\ 
\hline
76 & 195 & 223 & 75 & 39 & 10 & 2 & 0 & 0 & 0 & 0 & 0 & 0 & 0 & 0 & i & 3 & 0.1 & 5.57 & Hospital \\ 
\hline
61 & 171 & 274 & 96 & 50 & 12 & 2 & 0 & 0 & 0 & 0 & 0 & 0 & 0 & 0 & i & 3 & 0.15 & 5.91 & Hospital \\ 
\hline
48 & 134 & 275 & 134 & 63 & 19 & 4 & 0 & 0 & 0 & 0 & 0 & 0 & 0 & 0 & i & 3 & 0.2 & 6.51 & Hospital \\ 
\hline
69 & 96 & 83 & 27 & 5 & 1 & 0 & 0 & 0 & 0 & 0 & 0 & 0 & 0 & 0 & i & 4 & 0.01 & 4.4 & Hospital \\ 
\hline
72 & 98 & 83 & 27 & 5 & 1 & 0 & 0 & 0 & 0 & 0 & 0 & 0 & 0 & 0 & i & 4 & 0.05 & 4.33 & Hospital \\ 
\hline
63 & 116 & 85 & 27 & 5 & 1 & 0 & 0 & 0 & 0 & 0 & 0 & 0 & 0 & 0 & i & 4 & 0.1 & 4.43 & Hospital \\ 
\hline
53 & 120 & 109 & 29 & 5 & 1 & 0 & 0 & 0 & 0 & 0 & 0 & 0 & 0 & 0 & i & 4 & 0.15 & 4.64 & Hospital \\ 
\hline
42 & 97 & 132 & 35 & 7 & 1 & 0 & 0 & 0 & 0 & 0 & 0 & 0 & 0 & 0 & i & 4 & 0.2 & 4.92 & Hospital \\ 
\hline
48 & 47 & 32 & 2 & 0 & 0 & 0 & 0 & 0 & 0 & 0 & 0 & 0 & 0 & 0 & i & 5 & 0.01 & 3.63 & Hospital \\ 
\hline
49 & 47 & 32 & 2 & 0 & 0 & 0 & 0 & 0 & 0 & 0 & 0 & 0 & 0 & 0 & i & 5 & 0.05 & 3.56 & Hospital \\ 
\hline
49 & 51 & 32 & 2 & 0 & 0 & 0 & 0 & 0 & 0 & 0 & 0 & 0 & 0 & 0 & i & 5 & 0.1 & 3.5 & Hospital \\ 
\hline
43 & 59 & 32 & 2 & 0 & 0 & 0 & 0 & 0 & 0 & 0 & 0 & 0 & 0 & 0 & i & 5 & 0.15 & 3.58 & Hospital \\ 
\hline
37 & 58 & 42 & 2 & 0 & 0 & 0 & 0 & 0 & 0 & 0 & 0 & 0 & 0 & 0 & i & 5 & 0.2 & 3.73 & Hospital \\ 
\hline
95 & 71 & 212 & 274 & 421 & 769 & 1765 & 3632 & 5289 & 5382 & 3831 & 2054 & 839 & 308 & 42 & es & 3 & 0 & 1244.36 & Hospital \\ 
\hline
69 & 54 & 117 & 123 & 159 & 233 & 212 & 130 & 34 & 2 & 0 & 0 & 0 & 0 & 0 & es & 4 & 0 & 54.5 & Hospital \\ 
\hline
48 & 30 & 55 & 45 & 54 & 47 & 9 & 0 & 0 & 0 & 0 & 0 & 0 & 0 & 0 & es & 5 & 0 & 12.82 & Hospital \\ 
\hline
101 & 76 & 217 & 278 & 423 & 774 & 1770 & 3637 & 5293 & 5387 & 3834 & 2056 & 843 & 314 & 42 & is & 3 & 0.05 & 1389.16 & Hospital \\ 
\hline
72 & 57 & 122 & 124 & 160 & 234 & 214 & 132 & 35 & 2 & 0 & 0 & 0 & 0 & 0 & is & 4 & 0.05 & 640.38 & Hospital \\ 
\hline
49 & 31 & 57 & 46 & 55 & 48 & 9 & 0 & 0 & 0 & 0 & 0 & 0 & 0 & 0 & is & 5 & 0.05 & 22.39 & Hospital \\ 
\hline
658 & 15 & 0 & 0 & 0 & 0 & 0 & 0 & 0 & 0 & 0 & 0 & 0 & 0 & 0 & e & 3 & 0 & 2.4 & Highschool\_1 \\ 
\hline
325 & 8 & 0 & 0 & 0 & 0 & 0 & 0 & 0 & 0 & 0 & 0 & 0 & 0 & 0 & e & 4 & 0 & 2.44 & Highschool\_1 \\ 
\hline
151 & 1 & 0 & 0 & 0 & 0 & 0 & 0 & 0 & 0 & 0 & 0 & 0 & 0 & 0 & e & 5 & 0 & 2.35 & Highschool\_1 \\ 
\hline
78 & 0 & 0 & 0 & 0 & 0 & 0 & 0 & 0 & 0 & 0 & 0 & 0 & 0 & 0 & e & 6 & 0 & 2.42 & Highschool\_1 \\ 
\hline
20 & 0 & 0 & 0 & 0 & 0 & 0 & 0 & 0 & 0 & 0 & 0 & 0 & 0 & 0 & e & 7 & 0 & 2.41 & Highschool\_1 \\ 
\hline
658 & 11 & 0 & 0 & 0 & 0 & 0 & 0 & 0 & 0 & 0 & 0 & 0 & 0 & 0 & i & 3 & 0.01 & 2.98 & Highschool\_1 \\ 
\hline
659 & 11 & 0 & 0 & 0 & 0 & 0 & 0 & 0 & 0 & 0 & 0 & 0 & 0 & 0 & i & 3 & 0.05 & 2.95 & Highschool\_1 \\ 
\hline
671 & 11 & 0 & 0 & 0 & 0 & 0 & 0 & 0 & 0 & 0 & 0 & 0 & 0 & 0 & i & 3 & 0.1 & 2.89 & Highschool\_1 \\ 
\hline
673 & 11 & 0 & 0 & 0 & 0 & 0 & 0 & 0 & 0 & 0 & 0 & 0 & 0 & 0 & i & 3 & 0.15 & 2.88 & Highschool\_1 \\ 
\hline
673 & 11 & 0 & 0 & 0 & 0 & 0 & 0 & 0 & 0 & 0 & 0 & 0 & 0 & 0 & i & 3 & 0.2 & 2.84 & Highschool\_1 \\ 
\hline
325 & 5 & 0 & 0 & 0 & 0 & 0 & 0 & 0 & 0 & 0 & 0 & 0 & 0 & 0 & i & 4 & 0.01 & 3.03 & Highschool\_1 \\ 
\hline
325 & 5 & 0 & 0 & 0 & 0 & 0 & 0 & 0 & 0 & 0 & 0 & 0 & 0 & 0 & i & 4 & 0.05 & 2.97 & Highschool\_1 \\ 
\hline
329 & 5 & 0 & 0 & 0 & 0 & 0 & 0 & 0 & 0 & 0 & 0 & 0 & 0 & 0 & i & 4 & 0.1 & 2.92 & Highschool\_1 \\ 
\hline
331 & 5 & 0 & 0 & 0 & 0 & 0 & 0 & 0 & 0 & 0 & 0 & 0 & 0 & 0 & i & 4 & 0.15 & 2.84 & Highschool\_1 \\ 
\hline
330 & 5 & 0 & 0 & 0 & 0 & 0 & 0 & 0 & 0 & 0 & 0 & 0 & 0 & 0 & i & 4 & 0.2 & 2.83 & Highschool\_1 \\ 
\hline
151 & 1 & 0 & 0 & 0 & 0 & 0 & 0 & 0 & 0 & 0 & 0 & 0 & 0 & 0 & i & 5 & 0.01 & 3.02 & Highschool\_1 \\ 
\hline
151 & 1 & 0 & 0 & 0 & 0 & 0 & 0 & 0 & 0 & 0 & 0 & 0 & 0 & 0 & i & 5 & 0.05 & 2.95 & Highschool\_1 \\ 
\hline
152 & 1 & 0 & 0 & 0 & 0 & 0 & 0 & 0 & 0 & 0 & 0 & 0 & 0 & 0 & i & 5 & 0.1 & 2.89 & Highschool\_1 \\ 
\hline
156 & 1 & 0 & 0 & 0 & 0 & 0 & 0 & 0 & 0 & 0 & 0 & 0 & 0 & 0 & i & 5 & 0.15 & 2.9 & Highschool\_1 \\ 
\hline
157 & 1 & 0 & 0 & 0 & 0 & 0 & 0 & 0 & 0 & 0 & 0 & 0 & 0 & 0 & i & 5 & 0.2 & 2.86 & Highschool\_1 \\ 
\hline
78 & 0 & 0 & 0 & 0 & 0 & 0 & 0 & 0 & 0 & 0 & 0 & 0 & 0 & 0 & i & 6 & 0.01 & 2.97 & Highschool\_1 \\ 
\hline
78 & 0 & 0 & 0 & 0 & 0 & 0 & 0 & 0 & 0 & 0 & 0 & 0 & 0 & 0 & i & 6 & 0.05 & 2.95 & Highschool\_1 \\ 
\hline
78 & 0 & 0 & 0 & 0 & 0 & 0 & 0 & 0 & 0 & 0 & 0 & 0 & 0 & 0 & i & 6 & 0.1 & 2.91 & Highschool\_1 \\ 
\hline
80 & 0 & 0 & 0 & 0 & 0 & 0 & 0 & 0 & 0 & 0 & 0 & 0 & 0 & 0 & i & 6 & 0.15 & 2.9 & Highschool\_1 \\ 
\hline
82 & 0 & 0 & 0 & 0 & 0 & 0 & 0 & 0 & 0 & 0 & 0 & 0 & 0 & 0 & i & 6 & 0.2 & 2.85 & Highschool\_1 \\ 
\hline
20 & 0 & 0 & 0 & 0 & 0 & 0 & 0 & 0 & 0 & 0 & 0 & 0 & 0 & 0 & i & 7 & 0.01 & 2.97 & Highschool\_1 \\ 
\hline
20 & 0 & 0 & 0 & 0 & 0 & 0 & 0 & 0 & 0 & 0 & 0 & 0 & 0 & 0 & i & 7 & 0.05 & 2.94 & Highschool\_1 \\ 
\hline
20 & 0 & 0 & 0 & 0 & 0 & 0 & 0 & 0 & 0 & 0 & 0 & 0 & 0 & 0 & i & 7 & 0.1 & 2.93 & Highschool\_1 \\ 
\hline
20 & 0 & 0 & 0 & 0 & 0 & 0 & 0 & 0 & 0 & 0 & 0 & 0 & 0 & 0 & i & 7 & 0.15 & 2.9 & Highschool\_1 \\ 
\hline
21 & 0 & 0 & 0 & 0 & 0 & 0 & 0 & 0 & 0 & 0 & 0 & 0 & 0 & 0 & i & 7 & 0.2 & 2.86 & Highschool\_1 \\ 
\hline
658 & 23 & 0 & 0 & 0 & 0 & 0 & 0 & 0 & 0 & 0 & 0 & 0 & 0 & 0 & es & 3 & 0 & 2.34 & Highschool\_1 \\ 
\hline
325 & 8 & 0 & 0 & 0 & 0 & 0 & 0 & 0 & 0 & 0 & 0 & 0 & 0 & 0 & es & 4 & 0 & 2.4 & Highschool\_1 \\ 
\hline
151 & 2 & 0 & 0 & 0 & 0 & 0 & 0 & 0 & 0 & 0 & 0 & 0 & 0 & 0 & es & 5 & 0 & 2.4 & Highschool\_1 \\ 
\hline
78 & 1 & 0 & 0 & 0 & 0 & 0 & 0 & 0 & 0 & 0 & 0 & 0 & 0 & 0 & es & 6 & 0 & 2.33 & Highschool\_1 \\ 
\hline
20 & 0 & 0 & 0 & 0 & 0 & 0 & 0 & 0 & 0 & 0 & 0 & 0 & 0 & 0 & es & 7 & 0 & 2.37 & Highschool\_1 \\ 
\hline
659 & 23 & 0 & 0 & 0 & 0 & 0 & 0 & 0 & 0 & 0 & 0 & 0 & 0 & 0 & is & 3 & 0.05 & 3 & Highschool\_1 \\ 
\hline
325 & 8 & 0 & 0 & 0 & 0 & 0 & 0 & 0 & 0 & 0 & 0 & 0 & 0 & 0 & is & 4 & 0.05 & 2.93 & Highschool\_1 \\ 
\hline
151 & 2 & 0 & 0 & 0 & 0 & 0 & 0 & 0 & 0 & 0 & 0 & 0 & 0 & 0 & is & 5 & 0.05 & 2.93 & Highschool\_1 \\ 
\hline
78 & 1 & 0 & 0 & 0 & 0 & 0 & 0 & 0 & 0 & 0 & 0 & 0 & 0 & 0 & is & 6 & 0.05 & 2.93 & Highschool\_1 \\ 
\hline
20 & 0 & 0 & 0 & 0 & 0 & 0 & 0 & 0 & 0 & 0 & 0 & 0 & 0 & 0 & is & 7 & 0.05 & 2.92 & Highschool\_1 \\ 
\hline
117 & 180 & 140 & 75 & 38 & 16 & 0 & 0 & 0 & 0 & 0 & 0 & 0 & 0 & 0 & e & 3 & 0 & 3.32 & Highschool\_2 \\ 
\hline
99 & 115 & 77 & 38 & 13 & 2 & 0 & 0 & 0 & 0 & 0 & 0 & 0 & 0 & 0 & e & 4 & 0 & 2.99 & Highschool\_2 \\ 
\hline
76 & 78 & 41 & 20 & 0 & 0 & 0 & 0 & 0 & 0 & 0 & 0 & 0 & 0 & 0 & e & 5 & 0 & 2.91 & Highschool\_2 \\ 
\hline
59 & 51 & 30 & 9 & 0 & 0 & 0 & 0 & 0 & 0 & 0 & 0 & 0 & 0 & 0 & e & 6 & 0 & 2.76 & Highschool\_2 \\ 
\hline
44 & 31 & 17 & 2 & 0 & 0 & 0 & 0 & 0 & 0 & 0 & 0 & 0 & 0 & 0 & e & 7 & 0 & 2.64 & Highschool\_2 \\ 
\hline
117 & 164 & 71 & 26 & 7 & 3 & 0 & 0 & 0 & 0 & 0 & 0 & 0 & 0 & 0 & i & 3 & 0.01 & 3.32 & Highschool\_2 \\ 
\hline
117 & 171 & 71 & 26 & 7 & 3 & 0 & 0 & 0 & 0 & 0 & 0 & 0 & 0 & 0 & i & 3 & 0.05 & 3.32 & Highschool\_2 \\ 
\hline
92 & 196 & 91 & 27 & 7 & 3 & 0 & 0 & 0 & 0 & 0 & 0 & 0 & 0 & 0 & i & 3 & 0.1 & 3.31 & Highschool\_2 \\ 
\hline
75 & 169 & 113 & 30 & 7 & 3 & 0 & 0 & 0 & 0 & 0 & 0 & 0 & 0 & 0 & i & 3 & 0.15 & 3.34 & Highschool\_2 \\ 
\hline
63 & 140 & 110 & 28 & 4 & 2 & 0 & 0 & 0 & 0 & 0 & 0 & 0 & 0 & 0 & i & 3 & 0.2 & 3.24 & Highschool\_2 \\ 
\hline
99 & 101 & 37 & 8 & 4 & 0 & 0 & 0 & 0 & 0 & 0 & 0 & 0 & 0 & 0 & i & 4 & 0.01 & 4.34 & Highschool\_2 \\ 
\hline
99 & 103 & 37 & 8 & 4 & 0 & 0 & 0 & 0 & 0 & 0 & 0 & 0 & 0 & 0 & i & 4 & 0.05 & 3.17 & Highschool\_2 \\ 
\hline
78 & 127 & 39 & 8 & 4 & 0 & 0 & 0 & 0 & 0 & 0 & 0 & 0 & 0 & 0 & i & 4 & 0.1 & 3.12 & Highschool\_2 \\ 
\hline
67 & 130 & 56 & 8 & 4 & 0 & 0 & 0 & 0 & 0 & 0 & 0 & 0 & 0 & 0 & i & 4 & 0.15 & 3.17 & Highschool\_2 \\ 
\hline
55 & 109 & 56 & 4 & 2 & 1 & 0 & 0 & 0 & 0 & 0 & 0 & 0 & 0 & 0 & i & 4 & 0.2 & 3.09 & Highschool\_2 \\ 
\hline
76 & 66 & 18 & 3 & 0 & 0 & 0 & 0 & 0 & 0 & 0 & 0 & 0 & 0 & 0 & i & 5 & 0.01 & 3.08 & Highschool\_2 \\ 
\hline
79 & 66 & 18 & 3 & 0 & 0 & 0 & 0 & 0 & 0 & 0 & 0 & 0 & 0 & 0 & i & 5 & 0.05 & 3.06 & Highschool\_2 \\ 
\hline
69 & 79 & 19 & 3 & 0 & 0 & 0 & 0 & 0 & 0 & 0 & 0 & 0 & 0 & 0 & i & 5 & 0.1 & 3 & Highschool\_2 \\ 
\hline
56 & 93 & 22 & 3 & 0 & 0 & 0 & 0 & 0 & 0 & 0 & 0 & 0 & 0 & 0 & i & 5 & 0.15 & 3.04 & Highschool\_2 \\ 
\hline
46 & 87 & 29 & 3 & 0 & 0 & 0 & 0 & 0 & 0 & 0 & 0 & 0 & 0 & 0 & i & 5 & 0.2 & 3.03 & Highschool\_2 \\ 
\hline
59 & 43 & 12 & 3 & 0 & 0 & 0 & 0 & 0 & 0 & 0 & 0 & 0 & 0 & 0 & i & 6 & 0.01 & 3.07 & Highschool\_2 \\ 
\hline
61 & 43 & 12 & 3 & 0 & 0 & 0 & 0 & 0 & 0 & 0 & 0 & 0 & 0 & 0 & i & 6 & 0.05 & 3.01 & Highschool\_2 \\ 
\hline
58 & 51 & 12 & 3 & 0 & 0 & 0 & 0 & 0 & 0 & 0 & 0 & 0 & 0 & 0 & i & 6 & 0.1 & 3.02 & Highschool\_2 \\ 
\hline
47 & 61 & 13 & 3 & 0 & 0 & 0 & 0 & 0 & 0 & 0 & 0 & 0 & 0 & 0 & i & 6 & 0.15 & 3 & Highschool\_2 \\ 
\hline
40 & 65 & 14 & 3 & 0 & 0 & 0 & 0 & 0 & 0 & 0 & 0 & 0 & 0 & 0 & i & 6 & 0.2 & 3.02 & Highschool\_2 \\ 
\hline
44 & 24 & 7 & 1 & 0 & 0 & 0 & 0 & 0 & 0 & 0 & 0 & 0 & 0 & 0 & i & 7 & 0.01 & 3.01 & Highschool\_2 \\ 
\hline
45 & 24 & 7 & 1 & 0 & 0 & 0 & 0 & 0 & 0 & 0 & 0 & 0 & 0 & 0 & i & 7 & 0.05 & 2.96 & Highschool\_2 \\ 
\hline
44 & 24 & 7 & 1 & 0 & 0 & 0 & 0 & 0 & 0 & 0 & 0 & 0 & 0 & 0 & i & 7 & 0.1 & 2.9 & Highschool\_2 \\ 
\hline
38 & 25 & 7 & 1 & 0 & 0 & 0 & 0 & 0 & 0 & 0 & 0 & 0 & 0 & 0 & i & 7 & 0.15 & 2.96 & Highschool\_2 \\ 
\hline
34 & 29 & 7 & 1 & 0 & 0 & 0 & 0 & 0 & 0 & 0 & 0 & 0 & 0 & 0 & i & 7 & 0.2 & 2.91 & Highschool\_2 \\ 
\hline
117 & 79 & 155 & 139 & 170 & 190 & 126 & 66 & 34 & 12 & 0 & 0 & 0 & 0 & 0 & es & 3 & 0 & 12.95 & Highschool\_2 \\ 
\hline
99 & 65 & 92 & 73 & 73 & 43 & 24 & 7 & 1 & 0 & 0 & 0 & 0 & 0 & 0 & es & 4 & 0 & 7.69 & Highschool\_2 \\ 
\hline
76 & 48 & 55 & 37 & 31 & 17 & 7 & 0 & 0 & 0 & 0 & 0 & 0 & 0 & 0 & es & 5 & 0 & 4.97 & Highschool\_2 \\ 
\hline
59 & 33 & 35 & 20 & 11 & 2 & 0 & 0 & 0 & 0 & 0 & 0 & 0 & 0 & 0 & es & 6 & 0 & 3.99 & Highschool\_2 \\ 
\hline
44 & 22 & 18 & 9 & 1 & 0 & 0 & 0 & 0 & 0 & 0 & 0 & 0 & 0 & 0 & es & 7 & 0 & 3.05 & Highschool\_2 \\ 
\hline
117 & 84 & 163 & 147 & 176 & 192 & 126 & 66 & 34 & 12 & 0 & 0 & 0 & 0 & 0 & is & 3 & 0.05 & 16.55 & Highschool\_2 \\ 
\hline
99 & 67 & 95 & 77 & 75 & 43 & 24 & 7 & 1 & 0 & 0 & 0 & 0 & 0 & 0 & is & 4 & 0.05 & 7.83 & Highschool\_2 \\ 
\hline
79 & 49 & 57 & 39 & 31 & 17 & 7 & 0 & 0 & 0 & 0 & 0 & 0 & 0 & 0 & is & 5 & 0.05 & 5.66 & Highschool\_2 \\ 
\hline
61 & 34 & 36 & 21 & 11 & 2 & 0 & 0 & 0 & 0 & 0 & 0 & 0 & 0 & 0 & is & 6 & 0.05 & 4.6 & Highschool\_2 \\ 
\hline
45 & 23 & 19 & 9 & 1 & 0 & 0 & 0 & 0 & 0 & 0 & 0 & 0 & 0 & 0 & is & 7 & 0.05 & 3.63 & Highschool\_2 \\ 
\hline
31 & 11 & 2 & 0 & 0 & 0 & 0 & 0 & 0 & 0 & 0 & 0 & 0 & 0 & 0 & e & 100 & 0 & 38.58 & Sepsis \\ 
\hline
11 & 4 & 0 & 0 & 0 & 0 & 0 & 0 & 0 & 0 & 0 & 0 & 0 & 0 & 0 & e & 200 & 0 & 41.22 & Sepsis \\ 
\hline
1 & 0 & 0 & 0 & 0 & 0 & 0 & 0 & 0 & 0 & 0 & 0 & 0 & 0 & 0 & e & 500 & 0 & 39.73 & Sepsis \\ 
\hline
0 & 0 & 0 & 0 & 0 & 0 & 0 & 0 & 0 & 0 & 0 & 0 & 0 & 0 & 0 & e & 1000 & 0 & 42.24 & Sepsis \\ 
\hline
596 & 321 & 377 & 237 & 262 & 295 & 272 & 151 & 38 & 0 & 0 & 0 & 0 & 0 & 0 & i & 100 & 333 & 53.45 & Sepsis \\ 
\hline
645 & 329 & 286 & 129 & 109 & 74 & 41 & 16 & 5 & 0 & 0 & 0 & 0 & 0 & 0 & i & 100 & 666 & 50.46 & Sepsis \\ 
\hline
608 & 271 & 126 & 38 & 32 & 19 & 7 & 0 & 0 & 0 & 0 & 0 & 0 & 0 & 0 & i & 100 & 1333 & 46.7 & Sepsis \\ 
\hline
452 & 139 & 44 & 7 & 4 & 0 & 0 & 0 & 0 & 0 & 0 & 0 & 0 & 0 & 0 & i & 100 & 2666 & 48.32 & Sepsis \\ 
\hline
314 & 156 & 117 & 40 & 30 & 17 & 7 & 0 & 0 & 0 & 0 & 0 & 0 & 0 & 0 & i & 200 & 333 & 46.96 & Sepsis \\ 
\hline
282 & 129 & 51 & 9 & 4 & 0 & 0 & 0 & 0 & 0 & 0 & 0 & 0 & 0 & 0 & i & 200 & 666 & 49.41 & Sepsis \\ 
\hline
209 & 60 & 18 & 1 & 0 & 0 & 0 & 0 & 0 & 0 & 0 & 0 & 0 & 0 & 0 & i & 200 & 1333 & 45.36 & Sepsis \\ 
\hline
105 & 28 & 9 & 0 & 0 & 0 & 0 & 0 & 0 & 0 & 0 & 0 & 0 & 0 & 0 & i & 200 & 2666 & 50.28 & Sepsis \\ 
\hline
86 & 36 & 11 & 1 & 1 & 0 & 0 & 0 & 0 & 0 & 0 & 0 & 0 & 0 & 0 & i & 500 & 333 & 44.09 & Sepsis \\ 
\hline
58 & 17 & 3 & 0 & 0 & 0 & 0 & 0 & 0 & 0 & 0 & 0 & 0 & 0 & 0 & i & 500 & 666 & 45.92 & Sepsis \\ 
\hline
31 & 6 & 0 & 0 & 0 & 0 & 0 & 0 & 0 & 0 & 0 & 0 & 0 & 0 & 0 & i & 500 & 1333 & 48.91 & Sepsis \\ 
\hline
17 & 2 & 0 & 0 & 0 & 0 & 0 & 0 & 0 & 0 & 0 & 0 & 0 & 0 & 0 & i & 500 & 2666 & 47.98 & Sepsis \\ 
\hline
22 & 5 & 0 & 0 & 0 & 0 & 0 & 0 & 0 & 0 & 0 & 0 & 0 & 0 & 0 & i & 1000 & 333 & 46.78 & Sepsis \\ 
\hline
12 & 1 & 0 & 0 & 0 & 0 & 0 & 0 & 0 & 0 & 0 & 0 & 0 & 0 & 0 & i & 1000 & 666 & 44.35 & Sepsis \\ 
\hline
7 & 0 & 0 & 0 & 0 & 0 & 0 & 0 & 0 & 0 & 0 & 0 & 0 & 0 & 0 & i & 1000 & 1333 & 49.12 & Sepsis \\ 
\hline
1 & 0 & 0 & 0 & 0 & 0 & 0 & 0 & 0 & 0 & 0 & 0 & 0 & 0 & 0 & i & 1000 & 2666 & 48.68 & Sepsis \\ 
\hline
31 & 11 & 2 & 0 & 0 & 0 & 0 & 0 & 0 & 0 & 0 & 0 & 0 & 0 & 0 & es & 100 & 0 & 37.06 & Sepsis \\ 
\hline
11 & 4 & 0 & 0 & 0 & 0 & 0 & 0 & 0 & 0 & 0 & 0 & 0 & 0 & 0 & es & 200 & 0 & 41.21 & Sepsis \\ 
\hline
1 & 0 & 0 & 0 & 0 & 0 & 0 & 0 & 0 & 0 & 0 & 0 & 0 & 0 & 0 & es & 500 & 0 & 42.66 & Sepsis \\ 
\hline
0 & 0 & 0 & 0 & 0 & 0 & 0 & 0 & 0 & 0 & 0 & 0 & 0 & 0 & 0 & es & 1000 & 0 & 37.62 & Sepsis \\ 
\hline
645 & 336 & 411 & 265 & 358 & 499 & 701 & 756 & 417 & 0 & 0 & 0 & 0 & 0 & 0 & is & 100 & 666 & 61.45 & Sepsis \\ 
\hline
282 & 133 & 90 & 18 & 7 & 0 & 0 & 0 & 0 & 0 & 0 & 0 & 0 & 0 & 0 & is & 200 & 666 & 45.15 & Sepsis \\ 
\hline
58 & 17 & 6 & 0 & 0 & 0 & 0 & 0 & 0 & 0 & 0 & 0 & 0 & 0 & 0 & is & 500 & 666 & 46.22 & Sepsis \\ 
\hline
12 & 1 & 0 & 0 & 0 & 0 & 0 & 0 & 0 & 0 & 0 & 0 & 0 & 0 & 0 & is & 1000 & 666 & 46.68 & Sepsis \\
\hline
\end{longtable}
\end{scriptsize}
\end{landscape}
\bibliographystyle{IEEEtran}
\bibliography{references}